\documentclass[twocolumn,nofootinbib
]{revtex4-2}
\usepackage{hyperref}
\usepackage{amsmath}
\usepackage{bm}
\usepackage{amsfonts}
\usepackage{amssymb}
\usepackage{latexsym}
\usepackage{graphicx}
\usepackage[english]{babel}
\usepackage{multirow}
\usepackage{float}
\usepackage{url}
\usepackage{slashed}
\usepackage{xcolor} 
\usepackage[utf8]{inputenc}
\usepackage{physics}
\usepackage[export]{adjustbox}
\usepackage{verbatim}

\def\sfrac#1#2{{\textstyle{#1\over #2}}}
\newcommand{\be}{\begin{equation}}
	\newcommand{\ee}{\end{equation}}
\newcommand{\ba}{\begin{array}}
	\newcommand{\ea}{\end{array}}
\newcommand{\bea}{\begin{eqnarray}}
	\newcommand{\eea}{\end{eqnarray}}
\newcommand{\sss}{\scriptscriptstyle}

\newcommand{\nn}{\nonumber}

\newcommand{\N}{{\sss N}}
\renewcommand{\ss}{\mathrm{s}}

\begin{document}
\title{Dark Sector Electroweak Baryogenesis In Light Of The Galactic Center Excess}
\author{Jean-Samuel Roux}
\author{James M.\ Cline}
\affiliation{McGill University Department of Physics \& Trottier Space Institute, 3600 University St., Montr\'eal, QC H3A 2T8 Canada}
\begin{abstract}
   We revisit a model of electroweak baryogenesis that includes a dark matter candidate, and sequesters the new CP violation required to produce the baryon asymmetry in a dark sector.  The model can explain the baryon asymmetry, dark matter relic density, and the long-standing excess of gamma rays from the galactic center.  The first order electroweak phase transition induced by the new physics can give rise to gravitational waves that may be observed in future experiments.  The model predicts  dark matter signals in direct detectors,
   and a significant contribution to the Higgs boson invisible decay width.
\end{abstract}
\maketitle

\section{Introduction}

The particle physics nature of the dark matter of the Universe continues to be elusive, with so far no positive
signals from direct detection experiments or collider searches.  At the same time, a highly testable framework to understand the baryon asymmetry of the Universe, electroweak baryogenesis (EWBG), remains unconfirmed by collider searches for new particles at the TeV scale, that should be present to enable it.  However there is a long-standing hint for dark matter annihilation at the center of our galaxy, with a weak-scale cross section close to that needed for the ``WIMP miracle,'' which gives the desired dark matter relic density.
This suggests new physics at the weak scale, which could coincide with the requirements of EWBG.   These include the need for the electroweak phase transition to become first order, which might produce observable gravitational waves in several upcoming experiments.   
It is therefore timely to explore possible links between the 
galactic center excess (GCE) and EWBG.

The GCE is an excess of GeV-scale gamma rays coming from the galactic core,  detected by the Fermi Large Area Telescope \cite{Goodenough:2009gk}. The signal spectrum is diffuse and continuous, indicating that it is not the result of a decaying particle or the direct product of an annihilation process, which would produce monochromatic lines. Astrophysical sources, in particular millisecond pulsars (MSPs), have been proposed as a possible origin for the signal. However, the observed isotropy and smoothness of the GCE, and the lack of any resolved MSPs in the galactic center, challenge this explanation \cite{Hooper:2022bec}.  Many studies have attempted to disinguish the possible sources of the GCE 
by studying the statistics of the individual photons and the morphology of signal (cf. Refs.~\cite{Cholis:2021rpp,McDermott:2022zmq,Zhong:2024vyi} for a few recent examples), but so far they have yielded no definite preference between the DM and MSP hypotheses.
The Cherenkov Telescope Array is expected to provide a more definitive test within a few years \cite{Keith:2022xbd}.

A possible explanation for the GCE is the annihilation of a $\sim$ 50 GeV dark matter particle into standard model (SM) fermions. The annihilation cross section matches that of a thermal relic, $\ev{\sigma v}\sim10^{-26}$ cm$^3/$s. In particular, annihilations into bottom quarks \cite{Foster:2022nva,Cholis:2021rpp}, electrons, or muons \cite{DiMauro:2021qcf} provide a good fit to the GCE spectrum.

The $b\bar b$ annihilation channel  suggests 
models of Higgs portal DM, since $b\bar b$ would naturally be the dominant final state.
The Higgs portal is generic in models of electroweak baryogenesis (EWBG) that extend the SM with a gauge singlet real scalar field $s$.  These models typically introduce a tree-level barrier for the Higgs field which strengthens the 
electroweak phase transition, provided that $s$ gets a VEV.
Higgs-singlet mixing then leads to Higgs portal interactions with the dark sector.  

New sources of CP violation beyond the standard model are needed for EWBG, and these can often lead to severe constraints, due to improved limits on the electron and other
electric dipole moments.  This pressure can be eased by moving the CP violation into a hidden sector, which  
creates the opportunity to provide a dark matter candidate that plays a role in EWBG.

In the present work, we  revisit the model 
 proposed in Ref.\ \cite{Cline:2017qpe} in which the gauge singlet $s$ is coupled to an inert $SU(2)$ doublet, and to a $\sim 50$ GeV Majorana fermion, which is both a suitable dark matter candidate as well as a possible origin for the GCE.  We make a number of improvements on the original analysis.
 We fully take into account the explicit breaking of the $Z_2$ symmetry $s\to -s$, numerically solve for the nucleation action, including the full thermal potential and counterterms, and we use updated transport equations and determinations of the bubble wall velocity to predict the baryon asymmetry of the Universe (BAU). We quantify the production of gravitational waves (GWs) during the phase transition.  Finally, we show how the model is constrained by direct searches for dark matter, dipole moments, new charged particles, and invisible Higgs decays.

\section{Model}
To boost the strength of the electroweak phase transition (EWPT) from being a smooth crossover to first order, we adopt
the singlet scalar extension of the Standard Model, with explicit breaking of the $Z_2$ symmetry $s\to -s$.  Simplified treatments that impose this symmetry are possible, in which it is spontaneously broken in a first step leading to the EWPT, but then restored as the transition completes \cite{Espinosa:2011ax,Lewicki:2021pgr,Ellis:2022lft}.  In the present work, we are interested in the vacuum expectation value (VEV) of $s$ surviving at low temperatures, so that $s$ can mix with the Higgs boson at late times. $Z_2$ symmetry breaking is also required for EWBG; without it, the universe would be made of an equal number of regions with opposite baryon number, canceling the global BAU.
The full effective potential governing the phase transition can be written as 
\be 
V_{\rm eff} = V_0(h,s) + V_{CW}(h,s,T) + V_{T}(h,s,T) + \delta V(h,s),
\ee 
where the separate terms represent the tree-level, Coleman-Weinberg, thermal and counterterms potentials, respectively. 

In unitary gauge, the tree-level scalar potential is
\bea \label{eq:pot}
V_0 &=& -\sfrac{1}{2}\mu_h^2 h^2 + \sfrac{1}{4}\lambda_h h^4-\sfrac{1}{2}\mu_s^2s^2 + \sfrac{1}{4}\lambda_s s^4 + \sfrac{1}{4} \lambda_{hs} h^2s^2 \nn\\ &-& \sfrac{1}{2} A_{hs}h^2s - \sfrac{1}{3} A_3 s^3.
\label{V0eq}
\eea 
 The tadpole term $A_1^3 s$ can be removed without loss of generality by shifting the singlet field by a constant. We assume $A_{hs}>0$, which ensures $s$ acquires a positive VEV in the broken electroweak phase, as explained in section \ref{sec:FOPT}. 
 
 Setting the Higgs VEV to $\ev{h}\equiv v_0=246$ GeV and its mass to $m_h=125$ GeV allows us to fix the values of $\mu_h^2$ and $\lambda_h$, leaving $\mu_s^2, \lambda_s,\lambda_{hs},A_{hs}$ and $A_3$ as free parameters in the scalar potential. The other terms in the effective potential and the mass eigenvalues in our model are presented in Appendices \ref{app:masseigenvalues} and \ref{app:counterterms}.

Although the $Z_2$-breaking couplings $A_{hs}$ and $A_3$ could take any values, they are constrained by limits on Higgs-singlet mixing and stability of the electroweak symmetry breaking vacuum.  In our analytical estimates, we will therefore assume those parameters (which have dimensions of mass) are small compared to the weak scale. However we have imposed no strict bounds in our Monte Carlo numerical scan.

For EWBG, not only is a strong EWPT needed, but also a new source of CP violation that can produce particle asymmetries near the bubble walls.
For this purpose, and to introduce a dark matter candidate, 
 we include a Majorana fermion $\chi$ and an inert $SU(2)$ doublet $\phi$, which have the interactions
\be \label{eq:chi-int}
\mathcal{L} \supset \frac{1}{2} \bar{\chi}\big( M_\chi + \eta\, e^{i\gamma_5 \theta_\eta} s \big)\chi + \left[ y_\chi\bar{L}_\tau \phi P_R \chi + {\rm h.c.}\right]\,.
\ee 
The doublet $\phi$ has hypercharge $-1$, like SM left-handed fermions. We couple $\phi$ to a single SM lepton doublet $L_\tau=(\nu_\tau,\tau^-)^T$ to avoid constraints on lepton-flavor violation. 

For our parameters of interest, couplings to electrons or muons are generally ruled out by measurements of their electric and magnetic dipole moments as we discuss in Appendix~\ref{app:muon-electron}. 
We forbid those interactions by including an approximate $\tau$ lepton number global symmetry and by assigning tau lepton number 1 to $\phi$. Additionally, this excludes the lepton-number violating coupling $\lambda_5 (\phi^\dagger H)^2$, which could generate a large Majorana mass for $\nu_\tau$.  For simplicity, we have neglected tree-level lepton number-preserving couplings between $\phi$ and the SM Higgs doublet.

In Eq.\ (\ref{eq:chi-int}), we assume the mass $M_\chi$ and Yukawa coupling $y_\chi$ are real, since any complex phase can be absorbed by field redefinitions of $\chi$ and $\phi$. The
coupling $\eta$ is real by hermiticity,
and can be taken to be positive since a minus sign could be absorbed by  $\theta_\eta\to \theta_\eta + \pi$.   The $\theta_\eta$ phase is CP-violating if
it differs from $0$ or $\pi$; 
in that case, the singlet-fermion coupling is a combination of scalar and pseudoscalar interactions. During the EWPT, the singlet profile $s(z)$\footnote{$z$ denotes distance transverse to the bubble wall, with $z>0$ denoting the symmetric phase outside of the bubbles and $z<0$ the broken phase within.} induces a spatially-dependent $CP$-violating phase for the $\chi$ mass,
\be \label{eq:chi_phase}
\theta_{\chi}(z) = \arctan\left( \frac{\eta \sin\theta_\eta\,s(z)}{M_\chi + \eta \cos\theta_\eta \, s(z)}\right)\,,
\ee 
which gives rise to a CP-violating force that acts oppositely
on the two helicities of $\chi$.  The resulting helicity asymmetry near the bubble wall gets partially converted to a local $L_\tau$ asymmetry by inverse decays into $\phi$ and other scattering processes.
Finally, electroweak sphalerons redistribute the imbalance of left-handed leptons to baryons, resulting in the BAU.

\begin{figure}[t]
         \centering
        \includegraphics[scale=0.5,valign=c,page=2]{feyn-diagrams.pdf}\hspace{0.1\linewidth}
        \includegraphics[scale=0.5,valign=c,page=1]{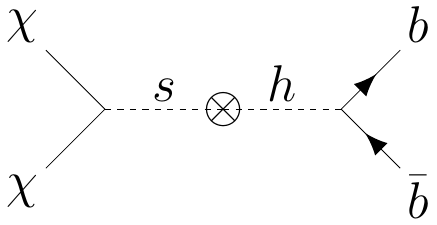}
    \caption{Main DM annihilation channels. The cross within a circle represents the Higgs-singlet mixing.}
    \label{fig:ann_diagrams}
\end{figure}

Because the singlet VEV does not vanish at zero temperature, the $\chi$ mass eigenvalue $m_\chi$ is different from the Lagrangian parameter $M_\chi$, cf.\ Eq.~(\ref{eq:chimass}). Here we focus on the case where $m_\chi< \{m_\phi,\ m_s\}$, making $\chi$ the dark matter candidate. Its stability is ensured by the
$Z_2$ symmetry under which $\chi$ and $\phi$ change sign.
Its relic density is set by the thermal freeze-out of the two annihilation channels shown in Fig.\ \ref{fig:ann_diagrams}, $\chi\chi\to L_\tau \bar{L}_\tau$ via $\phi$ exchange, and $\chi\chi\to b\bar{b}$ via Higgs portal coupling.\footnote{The Higgs portal allows decays to all SM fermions, but $b$ quarks
dominate.} The latter process can also explain the origin of the GCE signal, since DM annihilation into $b$ quarks has been shown to be compatible with the observed excess.

\section{First order EWPT}\label{sec:FOPT}
We begin by constructing the scalar potential and describing how the singlet field can make the EWPT
strongly first order, through a two-step phase transition in which $s$ first develops a VEV.  
 At zero temperature, there are two metastable vacua, which are located at
\be
\ev{h} = 0;\qquad  \ev{s} = \frac{A_3\pm\sqrt{A_3^2+4\lambda_s\mu_s^2}}{2\lambda_s}\approx \pm \sqrt{\frac{\mu_s^2}{\lambda_s}},
\ee
where we approximated $A_3\ll\lambda_s \mu_s$. In the true vacuum of the tree-level potential, the two scalar fields acquire VEVs $\ev{h}=v_0=246$ GeV and $\ev{s}=s_0$ that satisfy
\begin{align}
    \lambda_h v_0^2+\frac{1}{2}\lambda_{hs}s_0^2-A_{hs}s_0-\mu_h^2 &= 0,\label{muh}\\
    \lambda_s s_0^3+\frac{1}{2}\lambda_{hs}s_0v_0^2-\frac{1}{2}A_{hs}v_0^2-A_3 s_0^2-\mu_s^2s_0 &= 0.
\end{align}
For a given set of parameters, the latter equation can be solved for $s_0$ numerically or using Cardano's formula, but a good estimate is given by
\be 
s_0 \approx \frac{1}{2}\frac{A_{hs}v_0^2}{(\frac{1}{2}\lambda_{hs}v_0^2-\mu_s^2)} \approx \frac{1}{2}\frac{A_{hs}v_0^2}{m_{s}^2},\label{eq:w0approx}
\ee 
for small $Z_2$-breaking parameters. Assuming $A_{hs}>0$, $s_0$ is always positive. 

One can fix the value of $\mu_h^2$ using Eq.\ ($\ref{muh}$), then plug the result in the scalar mass matrix $\mathcal{M}^2$ to set $\lambda_h$ such that the Higgs mass eigenvalue is $m_h=125$ GeV, cf.\ Eq.~(\ref{eq:m2scalar}). Importantly, the gauge singlet and the Higgs boson mix after electroweak symmetry breaking, with mixing angle given by
\be 
\theta_{hs} =\frac{1}{2} \arctan\left(\frac{2\mathcal{M}_{hs}^2}{\mathcal{M}_{ss}^2-\mathcal{M}_{hh}^2}\right).
\ee 
 In the true vacuum  $\mathcal{M}_{hh}^2 = 2 \lambda_h v_0^2$, $\mathcal{M}_{ss}^2 = 2\lambda_s s_0^2-A_3 s_0+A_{hs}v_0^2/2s_0$ and $\mathcal{M}_{hs}^2=\lambda_{hs}v_0 s_0-A_{hs}v_0$.

At high temperature and in the limit of small $Z_2$-breaking terms, thermal corrections keep the fields in the electroweak symmetric ($h=0$) vacuum while $s$ gets a large VEV.\footnote{At very high temperature, the thermal potential keeps both fields near the origin, $(h,s)=(0,0)$. The singlet scalar gets a VEV when the temperature drops below another critical point, $T_c'\approx \mu_s/\sqrt{c_s}$, where $c_s$ is defined below Eq.~(\ref{eq:highT}). We require $T_c'\gg T_c$.} Below a critical temperature $T_c$, this vacuum becomes metastable and, assuming there is a barrier in the potential, the fields can tunnel to the true global minimum at $h\neq0$. The tunneling rate per unit volume $\Gamma$ is set by the $O(3)$-symmetric Euclidean bubble action $S_3$,
\be \label{eq:nucRate}
\Gamma\approx T^4\left(\frac{S_3}{2 \pi T}\right)^{3 / 2} \exp \left(-\frac{S_3}{T}\right),
\ee 
where 
\be \label{eq:action}
S_3=4 \pi \int r^2 d r\left(\frac{1}{2}\left(\frac{d h}{d r}\right)^2+\frac{1}{2}\left(\frac{d s}{d r}\right)^2+V_{\mathrm{eff}}\right).
\ee 

The relevant temperature for the phase transition is the nucleation temperature $T_n$ at which the probability that a bubble nucleated inside a Hubble volume reaches unity \cite{Moreno:1998bq},
\be \label{eq:Tnucdef}
1=\int_{T_n}^{T_c}\frac{dT}{T}\frac{\Gamma}{H^4}=\int_{T_n}^{T_c}dT\frac{T^3}{H^4}\left(\frac{S_3}{2\pi T}\right)^{3/2}e^{-S_3/T}.
\ee 
 For the EWPT, the nucleation temperature satisfies $S_3/T_n\approx 140$. The strength of the phase transition is characterized by the ratio $v_n/T_n$, where $v_n$ is the Higgs VEV at $T_n$. Preventing washout of the baryon asymmetry by electroweak sphalerons inside the bubble requires $v_n/T_n\gtrsim 1.1$.

\subsection{Impact of $Z_2$ breaking parameters}
\label{Z2sect}
Breaking the $Z_2$ symmetry $s\to -s$ is essential if one wants to achieve EWBG and study Higgs-singlet mixing at low temperatures.  In Ref.\ \cite{Cline:2017qpe}, it was sufficient to posit a very small degree of $Z_2$ breaking for EWBG, but larger levels have interesting effects on the EWPT,
as we summarize here \cite{Profumo:2007wc,Profumo:2014opa,Hashino:2016xoj,Balazs:2016tbi,Beniwal:2018hyi,Harigaya:2022ptp}.

First, as mentioned above, the trilinear coupling $A_{hs}$ controls the singlet VEV in the true vaccuum of the theory ($s_0\propto A_{hs}$), which in turn directly impacts the mixing angle between the scalar fields ($\mathcal{M}^2_{hs}\propto A_{hs}$) and thus the coupling of $\chi$ and $s$ to SM fermions. 
But the breaking of $Z_2$ can also enhance the strength of the phase transition as we now review.
In addition to the two couplings $A_{hs}$ and $A_3$ that appear in the tree-level scalar potential (\ref{eq:pot}), thermal corrections induce a third parameter $A_T$ that breaks $Z_2$, as can be seen in the high-$T$ limit,
\be \label{eq:highT}
V(h,s,T) \approx  V_0(h,s) +\frac{1}{2}c_h T^2h^2+ \frac{1}{2} c_s T^2 s^2  - A_T  T^2 s,
\ee 
where $V_0(h,s)$ is the zero-temperature tree-level potential and
\begin{align*}
c_h &= \frac{1}{48} (9 g^2 + 3 g'^2 + 12 y_t^2 + 24 \lambda_h + 2 \lambda_{hs}),\\
c_s &= \frac{1}{12} (2 \lambda_{hs} + 3 \lambda_s + \eta^2),\\
A_T &=\frac{1}{12} ( A_3 + 2 A_{hs} -  M_\chi \eta \cos{\theta_\eta}).
\end{align*}
(Recall that $M_\chi$ is the bare mass term defined in Eq.~(\ref{eq:chi-int})). Here we only keep terms that are quadratic in $T$, and we ignore the CW potential as well as daisy resummation. 
Instead, we work to linear order in $Z_2$-breaking parameters $A\equiv\{A_{hs},A_{3},A_T\}$. The full one-loop potential including resummation is difficult to analyze analytically, but it would only slightly shift the parameter space where the following results apply, and the argument given here should be valid for a wide range of models.

Although the strength of the phase transition is properly expressed by the ratio $v_n/T_n$ at nucleation, analytical treatment of the nucleation temperature and its dependence on $A$ parameters via Eq.~(\ref{eq:Tnucdef}) is laborious. We will instead consider the impact of $A$ on the critical temperature $T_c$ and corresponding Higgs VEV $v_c$, keeping in mind that $v_n/T_n$ is strongly correlated with $v_c/T_c$ \cite{Vaskonen:2016yiu,Cline:2017qpe}. 

In what follows, parameters denoted with a tilde will refer to the expected values in the limit of $Z_2$ symmetry, $A\to 0$. In the electroweak symmetric phase ($h=0$), there are two minima at
\begin{align}
s(T) &= \pm \sqrt{\frac{\mu_s^2-c_s T^2}{\lambda_s}} + \frac{1}{2}\left(\frac{A_3}{\lambda_s} + \frac{A_T T^2}{\mu_s^2-c_s T^2}\right) + \mathcal{O}(A^2)\\&\equiv \pm \tilde{s}(T) + \frac{1}{2 \lambda_s} \left(A_3 + A_T\frac{T^2 }{\tilde{s}^2}\right), \nn
\end{align}
where $\tilde{s}=\sqrt{(\mu_s^2-c_s T^2)/\lambda_s}$. The $A$ parameters break the degeneracy between the two minima, such that the potential of the global minimum is lower than the $Z_2$ symmetric case,
\begin{align}\label{eq:vst}
V_S(T) &= -\frac{(\mu_s^2-c_s T^2)^2}{4\lambda_s} - \tilde{s}^3\abs{\frac{A_3}{3} + A_T\frac{T^2}{\tilde{s}^2}}+ \mathcal{O}(A^2)\\
& \equiv \tilde{V}_S(T) - \tilde{s}^3\mathcal{A}(T).\nn
\end{align}
Here we introduced the linear combination $\mathcal{A}(T)=\abs{A_3/3 + A_T T^2/\tilde{s}^2}$.

On the other hand, in the broken electroweak phase ($h\neq0$), the singlet acquires an $\mathcal{O}(A)$ VEV, cf.\ Eq.~(\ref{eq:w0approx}), which has no impact on the potential energy to linear order in $A$. Therefore, the Higgs VEV and potential energy at a given temperature are equal to those in the $Z_2$-symmetric scenario,
\begin{align} 
h(T) &= \tilde{v}(T)\equiv \sqrt{\frac{\mu_h^2-c_h T^2}{\lambda_h}}, \label{eq:vc}\\
\label{eq:vbt}
V_B(T)&= \tilde{V}_B(T) \equiv -\frac{(\mu_h^2-c_h T^2)^2}{4\lambda_h}.
\end{align}
 
Eqs.\ (\ref{eq:vst}) and (\ref{eq:vbt}) show that $\mathcal{A}(T)$ decreases the potential difference between the false and true vacuum (\textit{i.e.} $\Delta V=V_S-V_B$ is more negative at high temperatures). This lowers the critical temperature $T_c$ (and the nucleation temperature $T_n$, by extension) as it takes longer for the potential of the broken phase to become degenerate with the electroweak symmetric vacuum. This in turn increases the Higgs VEV at $T_c$ by virtue of Eq.\ (\ref{eq:vc}) and increases the ratio $v_c/T_c$, resulting in a stronger phase transition.

We can find the shift in critical temperature by equating $V_B(T)$ and $V_S(T)$, making the ansatz $T_c^2=\tilde{T}_c^2+\delta T^2$ and Taylor expanding to linear order in $\delta T^2$. We find
\be\label{eq:dT}
\delta T^2 = -\frac{\tilde{s}_c^3 \mathcal{A}_c }{|\tilde{V}_c|^{1/2}\left(c_h/\sqrt{\lambda_h}-c_s/\sqrt{\lambda_s}\right)}.
\ee 
The subscript $c$ indicates quantities evaluated at the $Z_2$-symmetric critical temperature, e.g. $\mathcal{A}_c=\mathcal{A}(\tilde{T}_c)$, where 
\be 
\tilde{T}_c^2 = \frac{\mu_h^2/\sqrt{\lambda_h} - \mu_s^2/\sqrt{\lambda_s}}{c_h/ \sqrt{\lambda_h}-c_s/\sqrt{\lambda_s}}.
\ee 
In Eq.~(\ref{eq:dT}), $\tilde{V}_c$ is the potential energy of both minima at $\tilde{T}_c$, in the limit $A\to 0$. Note that $\delta T^2$ is always negative since $c_h/\sqrt{\lambda_h} >c_s/\sqrt{\lambda_s}$ is a necessary condition in the $Z_2$-symmetric model for the broken vacuum $h\neq 0$ to be the global minimum below the critical temperature. Therefore, explicit $Z_2$-breaking terms always tend to lower the critical temperature and increase the strength of the phase transition.
We have checked this by performing a numerical scan where all parameters were fixed except for $A_3$, $A_{hs}$ and $\cos{\theta_\eta}$, which were chosen randomly in the following uniform distributions,
\begin{align} 
A_3 & \in [-5,5 ]\ \mathrm{GeV},\\
A_{hs} & \in [0,5]\ \mathrm{GeV},\\
\cos{\theta_\eta} & \in \ [-0.2,0.2].
\end{align} 
Other parameters were fixed to those of Model 2 in Table \ref{tab1}.\footnote{In our scan, we find that the analytical expressions for $\tilde{s}$, $\tilde{V}_c$ and $\tilde{T}_c$ are generally in poor agreement with numerical values obtained with CosmoTransitions. This is because our argument ignored contributions such as the CW potential and daisy resummation, which do not change our conclusion but slightly shift the critical temperature and fields VEVs when taken into account. For this reason we used the numerical values of $\tilde{s}$, $\tilde{V}_c$ and $\tilde{T}_c$ rather than the analytical expressions in our estimate of $\delta T_c^2$.}

\begin{figure}[t]
    \centering
    \includegraphics[width=\linewidth]{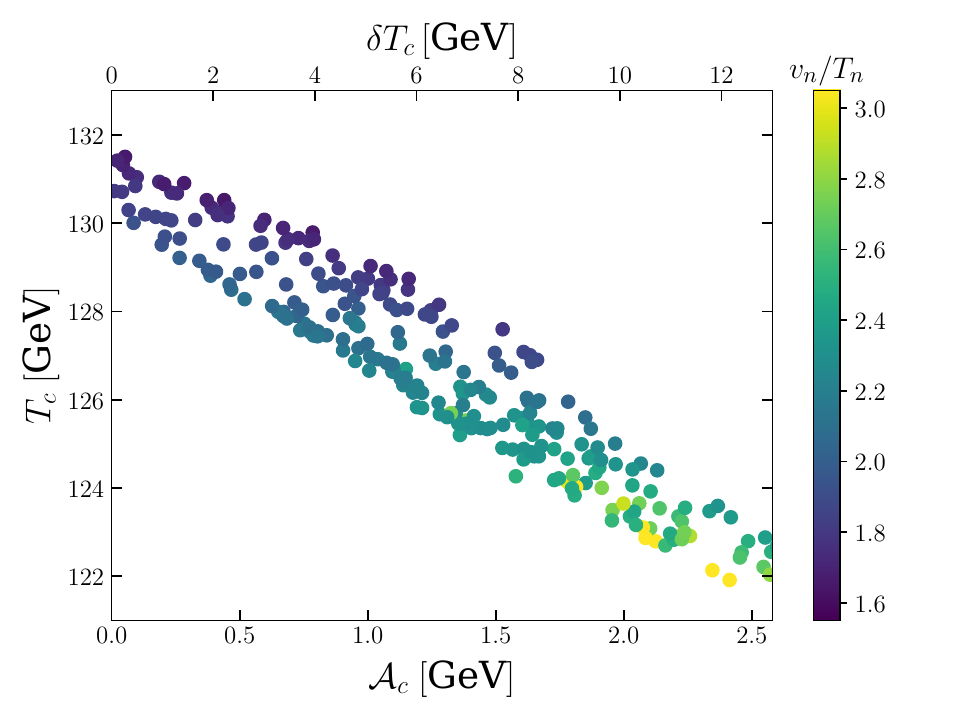}
    \caption{Variation of the critical temperature $T_c$ with the linear combination $\mathcal{A}_c$ defined in the text for our numerical scan, where $Z_2$ breaking parameters $A_3$, $A_{hs}$ and $\cos{\theta_\eta}$ where varied randomly. The upper axis shows the corresponding values of estimated temperature shift $\delta T_c$ obtained from Eq.~(\ref{eq:dT}). Coloring of the dots indicates the strength of the phase transition $v_n/T_n$. }
    \label{fig:Tc}
\end{figure}

In Fig.~\ref{fig:Tc}, we plot the critical temperature as a function of $\mathcal{A}_c$ (lower axis) and $\delta T_c\equiv \delta T_c^2/(2 \tilde{T}_c)$  (upper axis) evaluated using Eq.~(\ref{eq:dT}). As expected, the critical temperature decreases linearly with the $Z_2$-breaking parameters, and the strength of the phase transition increases, as indicated by the color of the dots. With the parameters we have selected, $Z_2$-breaking parameters of even a few GeV are enough to nearly double the strength of the phase transition.

We also note that explicit $Z_2$-breaking terms favor GW production, as will be discussed in detail in the next Section. In particular, the lower nucleation temperature implies that the latent heat $\alpha \sim \Delta V/\rho_{\mathrm{rad}}$ must be evaluated at a lower temperature, and even though $\Delta V$ decreases with $\mathcal{A}(T)$, the denominator scales as $\sim T_n^4$ and decreases more rapidly. Therefore, the EWPT releases more energy than in the $Z_2$-symmetric scenario.

\begin{figure*}[t]
    \centering
    \includegraphics[width=0.49\linewidth]{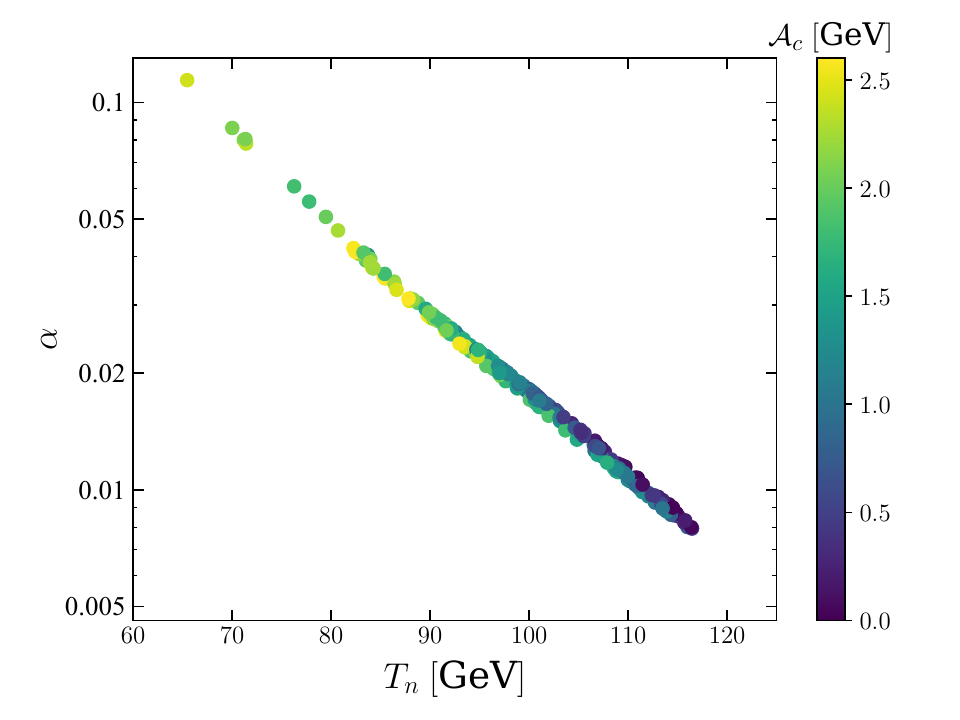}
    \includegraphics[width=0.49\linewidth]{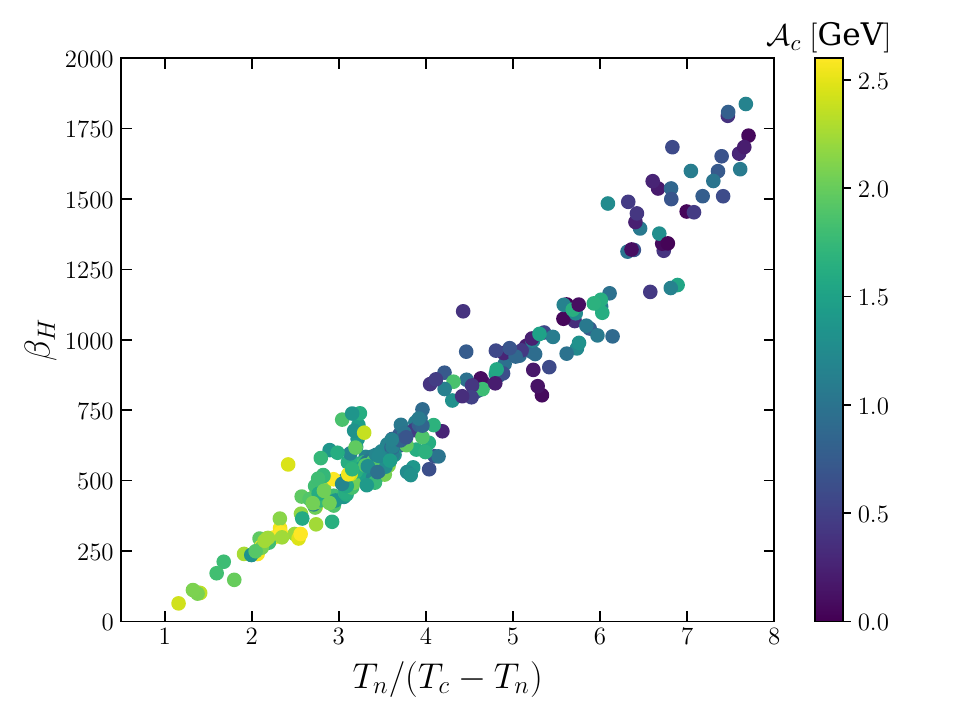}
    \caption{Correlation of $\alpha$ (left) and $\beta_H$ (right) with the critical and nucleation temperatures in our second scan. Coloring of the dots indicates the temperature shift parameter $\delta T_c$ due to explicit $Z_2$ breaking in each model.}
    \label{fig:alpha-beta-Tc}
\end{figure*}
The nucleation rate $\beta_H$ is also impacted by the lower nucleation temperature, as we can estimate from Taylor expanding the bounce action about $T_n$ as in Ref.~\cite{Anderson:1991zb},
\be 
\frac{S_3(T)}{T} = \frac{S_3(T_n)}{T_n}\left(1+2x+3x^2+\mathcal{O}(x^3)\right),
\ee
where $x=(T-T_n)/(T_c-T_n)$.  This expansion is valid for $T_n\lesssim T_c$. We also recall that $S_3(T_n)/T_n\approx 140$. One can then apply the definition of $\beta_H$, Eq.~(\ref{eq:beta}), to find that it approximately scales as $T_n/(T_c-T_n)$. $Z_2$-breaking parameters decrease the nucleation temperature and increase the difference $T_c-T_n$, thus making $\beta_H$ smaller. Fig.~\ref{fig:alpha-beta-Tc} shows the variation of $\alpha$ and $\beta_H$ in our scan and their correlations with $\mathcal{A}_c$.

\section{Gravitational waves}\label{sec:gw}
Sound waves propagating in the plasma as a result of the FOPT are the main source of gravitational waves in our model. To estimate the strength of the GW signal and compare it with the sensitivity of future detectors, one must compute the gravitational wave spectrum $\Omega_{\mathrm{gw}}$ from the dynamical parameters of the phase transition. 

There are three parameters that are needed to quantify GW production.
(i)~The vacuum energy released in the plasma during the FOPT, relative to the radiation energy density in the plasma, is
\be 
\alpha=\frac1{\rho_\gamma}\left(\Delta V-\frac{T_n}4\Delta\frac{dV}{dT}\right),
\ee 
where $\rho_\gamma = \pi^2 g_* T_n^4/30$ and $\Delta$ indicates the difference between the false and true vacuum of the transition.
(ii)~The inverse timescale of the phase transition is characterized by the parameter
\be \label{eq:beta}
\beta_H=\frac{\beta}{H}=T_n\left.\frac{d}{dT}\left(\frac{S_3}{T}\right)\right|_{T=T_n}.
\ee 
GW production is typically favored for large $\alpha$ and small $\beta_H$.
(iii)~The final parameter is the wall velocity $v_w$, which also plays a significant role in the computation of the BAU. When our model was first proposed in Ref.\ \cite{Cline:2017qpe}, the wall was assumed to be subsonic, $v_w\sim 0.1$. However, many recent papers have shown that $v_w$ is usually closer to the speed of sound in the plasma, $v_w\sim c_s=1/\sqrt{3}$, for singlet scalar extensions of the SM. Very strong transitions ($\alpha \gtrsim 0.1$) can even lead to ultra-relativistic velocities, $v_w\to 1$. The latter scenario is undesirable in the context of EWBG, as it leads to vanishing BAU.

Determination of the wall velocity is a difficult problem,  requiring the solution of Boltzmann equations for fluid perturbations, coupled to the scalar field equations of motion. Their solutions were considered by several studies in the context of the EWPT within the singlet scalar extension of the SM, using various approximations (cf.\ \cite{Cline:2021iff,Lewicki:2021pgr,Laurent:2022jrs,Ellis:2022lft} for a few recent examples). For instance, Ref.\ \cite{Laurent:2022jrs} found that transitions of moderate strength ($0.01\lesssim \alpha\lesssim0.04$) often, but not always, lead to runaway detonation solutions ($v_w\to1$); whereas Ref.\ \cite{Ellis:2022lft} performed an analysis in the non-$Z_2$-symmetric model and found that, in the same range of $\alpha$, the wall velocity fell in a hybrid regime $c_s\lesssim v_w\lesssim v_J$, where $v_J$ is the Chapman-Jouguet velocity,
\be 
v_J=\frac{1}{\sqrt{3}}\frac{1+\sqrt{3\alpha^2+2\alpha}}{1+\alpha}\,.
\ee 
The disparity between these results is likely due to the different regions of parameter space scanned in each study. For simplicity, we will follow the results of the non-$Z_2$-symmetric scenario and assume $v_w\approx v_J$, keeping in mind that a more accurate treatment could be necessary to correctly assess the observability of GWs and the success of the EWBG.

The observable GW spectrum produced by sound waves in the plasma is \cite{Hindmarsh:2020hop}
\be 
\Omega_{\mathrm{gw}}(f)h^2=2.06\,F_{\mathrm{gw},0}\,\tilde{\Omega}_{\mathrm{gw}}\frac{(H_{*}R_{*})^{2}}{\sqrt{K}+H_{*}R_{*}}K^{2}C(f/f_{\mathrm{p},0}),
\ee 
where $F_{\mathrm{gw,0}}=3.57\times 10^{-5}\,(100/g_*)^{1/3}$ is the energy redshift due to the expansion of the universe, $\tilde{\Omega}_{\mathrm{gw}}=0.012$ is a numerically determined constant that measures the efficiency of GW production, $H_*$ is the Hubble parameter at percolation, and $R_*$ is the mean bubble center separation, given by $H_*R_*\approx (8\pi)^{1/3} v_w/\beta_H$. The kinetic energy fraction is $K=\kappa\alpha/(1+\alpha)$ where, assuming $v_w\approx v_J$, $\kappa\simeq \sqrt{\alpha}/[0.135+\sqrt{0.98+\alpha}]$  is the efficiency with which vacuum energy is converted into kinetic energy \cite{Espinosa:2010hh}. The last factor is $C(s)=s^3\big(7/(4+3s^2)\big)^{7/2}$, the spectral shape around the peak frequency,
\be 
f_{p,0}\simeq2.62\times 10^{-5}\left(\frac{1}{H_*R_*}\right)\left(\frac{T_\text{n}}{10^2\ \text{GeV}}\right)\left(\frac{g_{*}}{100}\right)^{\frac{1}{6}}\ \text{Hz}.
\ee 

Having determined the spectrum, we can check whether a GW signal is detectable by computing the signal-to-noise ratio (SNR),
\be \label{eq:snr}
\mathrm{SNR}=\sqrt{\mathcal{T}\int_{f_{\min}}^{f_{\max}}df\left[\frac{\Omega_{\mathrm{gw}}(f)}{\Omega_{\mathrm{noise}}(f)}\right]^2}.
\ee 
Here, $\mathcal{T}$ is the duration of the experiment and $\Omega_{\mathrm{noise}}$ is the noise spectrum of a given detector. Whenever the SNR is greater than a certain threshold SNR$_{\mathrm{thr}}$ the signal is detectable. We take $\mathcal{T}=4$ years and SNR$_{\mathrm{thr}}=10$.

\begin{table}[]
\centering
\begin{tabular}{ | c | c | c | c | c | c|c|c|c|c|c|} 
 \hline
 Model & $\mu_s$ & $\lambda_s$ &  $\lambda_{hs}$ & $A_{hs}$ & $A_3$ & $M_\chi$ & $\eta$ & $\theta_\eta$ & $y_\chi$ &$m_\phi$\\\hline

 1 & 58.6& 0.12&0.42 &2.4 &-5.4 & 48.6& 0.42&-1.4 &0.61 &132.8\\
2 & 60.0& 0.10 &0.48 &0.7 &-2.6 &63.8 & 0.21& 0.21&0.56&101.9\\\hline
Mean & 62.2 & 0.20&0.92 & 7.4 & 1.5  & 59.6 & 0.74& -0.14& 0.66& 127.3\\ 
Std. dev. & 9.6 & 0.11 &0.38& 8.5 &10.4&22.0 & 0.32&1.9& 0.10& 15.0\\
\hline
\end{tabular}
\caption{Lagrangian parameters for two benchmark models and statistical distribution of models that yield successful baryogenesis and DM abundance, as explained in the text.  Dimensionful quantities are in GeV. $M_\chi$ is the bare mass term appearing in Eq.~(\ref{eq:chi-int}); the  mass eigenvalue $m_\chi$ is field dependent (cf.\ Eq.~(\ref{eq:chimass})) and given in Table~\ref{tab2}. In both models, $m_s\sim 100\,$GeV (see Table \ref{tab2}).}
\label{tab1}
\end{table}

\begin{figure}[t]
    \centering
    \includegraphics[width=\linewidth]{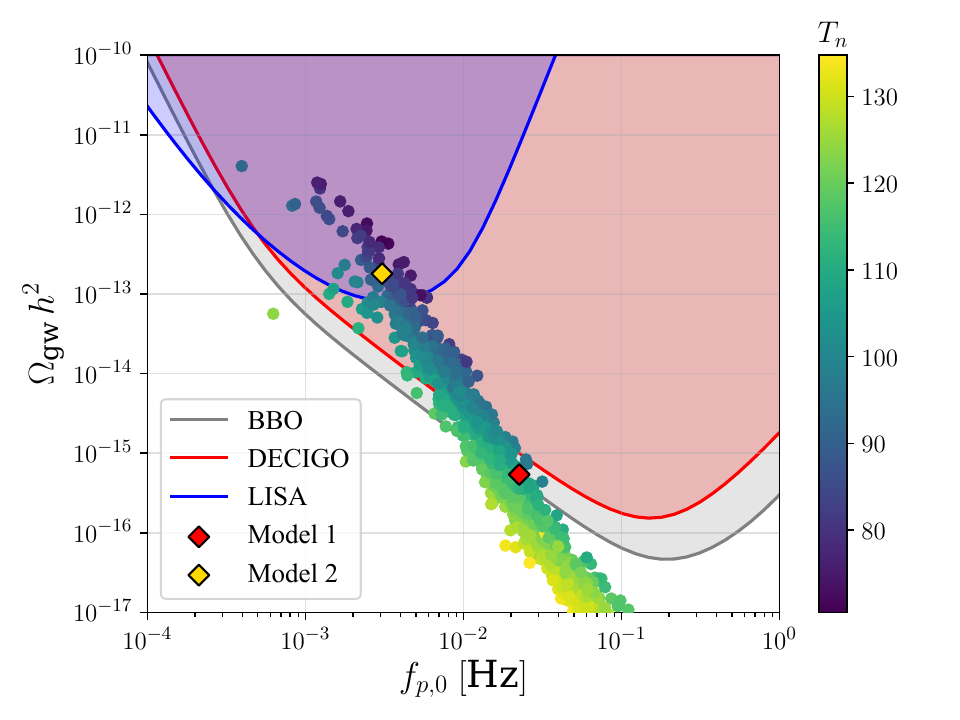}
    \caption{Comparison of the peak amplitude and frequency of the gravitational wave spectrum with peak-integrated sensitivity curves of future detectors. Coloring of the dots shows the nucleation temperature $T_n$. Also shown are the two benchmark models of Table \ref{tab1}.}
    \label{fig:gw}
\end{figure}

Instead of computing the integral of Eq.~(\ref{eq:snr}) directly, we use the peak-integrated sensitivity curves (PISCs) introduced in Refs. \cite{Alanne:2019bsm,Schmitz:2020syl} to determine whether a given GW spectrum is observable. Normalization of the PISCs is such that, whenever the peak amplitude is greater than $\Omega_{\mathrm{PISC}}h^2$, the SNR is above the detection threshold. We use the fit functions given in Ref.\ \cite{Schmitz:2020syl} for LISA, DECIGO and BBO.
These curves are shown in Fig.\ \ref{fig:gw}, along with prediction from our main MCMC scan over parameters, to be described in Section \ref{sect:results}.  One can see that models with a range of sufficiently low nucleation temperatures
$T_n \lesssim 120\,$GeV are likely to be detectable.
(Figs.\ \ref{fig:Tc} and\ \ref{fig:alpha-beta-Tc} show results from the separate scan over $Z_2$-breaking parameters of Section \ref{Z2sect},
which explored a more limited range of nucleation temperatures.)

\begin{figure}[t]
    \centering
    \includegraphics[width=\linewidth]{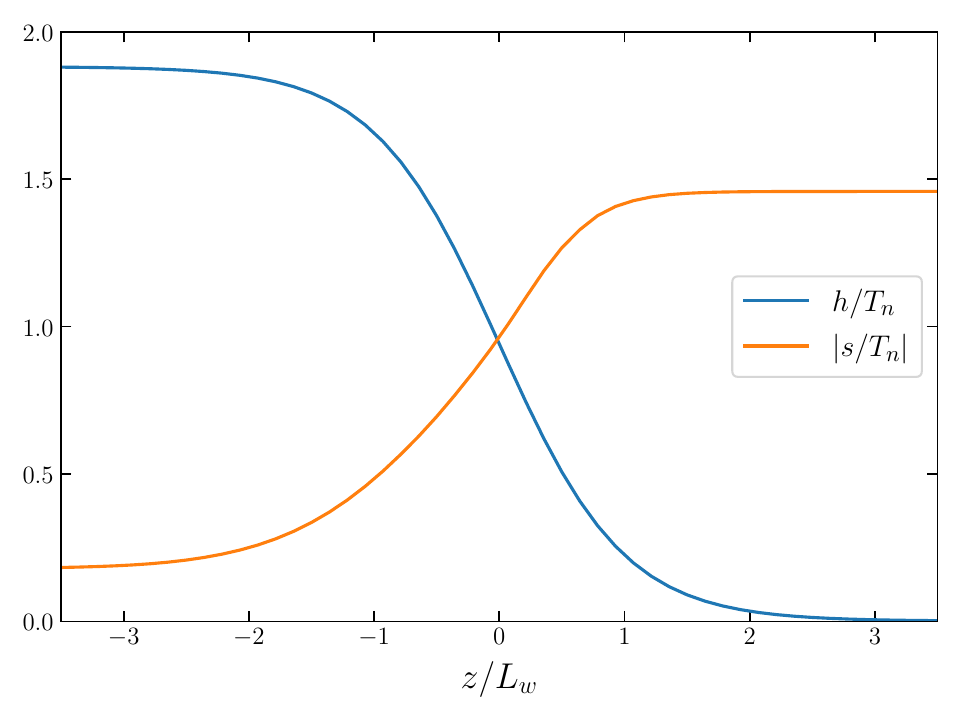}
    \caption{Example of wall profiles $h(z)$ and $s(z)$ at nucleation for Model 1 of Table \ref{tab1}.}
    \label{fig:profile}
\end{figure}
\section{Wall profiles}
Knowledge of the bubble wall profile $(h(z),s(z))$ is essential to solve the transport equation and evaluate the resulting BAU. Its full computation involves solving a system of partial differential equations for the scalar fields coupled to a perfect background fluid, which is numerically expensive and outside the scope of this work. Ref.\ \cite{Friedlander:2020tnq} proposed a self-consistent algorithm to obtain the wall shape in strong EWPTs with a scalar singlet and found that the fields could be described by a tanh profile, to a good approximation. For simplicity, we therefore assume the Higgs field follows the ansatz
\be \label{eq:hz}
h(z) = {v_n \over 2}\left(1-\tanh\left(z\over L_w\right)\right),
\ee 
where $L_w$ is the wall thickness. 

A similar ansatz is often used for the singlet profile. Here, we instead follow the strategy used in Ref.\ \cite{Cline:2017qpe} by minimizing the potential along the radial coordinate in field space, $(\partial_\rho V_{\rm eff})|_\theta=0$, for all values of $\theta$ between the two vacua. Combining the resulting field space trajectory with the spatial dependence of Eq.~(\ref{eq:hz}) allows us to get the singlet wall profile. The wall profiles for Model 1 of Table \ref{tab1} are illustrated in Fig.\ \ref{fig:profile} (the benchmark models will be described in Section \ref{sect:results}).  For numerical purposes, we found that $s(z)$ can be well fit by a deformed hyperbolic tangent function, 
\be 
s(z)\approx s_{0n}+ \frac{s_n-s_{0n}}{2}\left[1+\tanh\bigg(P_5\left(z/ L_w\right)\bigg)\right],
\ee
where $P_5(x)=\sum_{i=0}^5 a_i\, x^i$ is a fifth-order polynomial and the coefficients $a_i$ are parameters of the fit.

 We estimate $L_w$ following Ref.~\cite{Espinosa:2011eu}. In the thin-wall approximation, the bubble radius $R$ is much bigger than the wall thickness. One can therefore substitute $r\to R$ and $dr\to dz$ in the bubble action (\ref{eq:action}), where $R$ is a constant and $z$ is the direction transverse to the wall. Assuming the false and true vacua are nearly degenerate, the action can then be approximated by
 \be 
S_3 \approx {c_1 \over L_w} (v_n^2+\Delta s_n^2) + c_2 L_w V_b\,,
 \ee 
 where $c_1,\ c_2$ are numerical constants, $\Delta s_n=s_{0n}-s_n$ and $V_b$ is the potential barrier separating the two minima, which we evaluated numerically. Minimizing $S_3$ with respect to $L_w$ leads to 
 \be 
L_w^2 \approx {c_1\over c_2} \frac{v_n^2+\Delta s_n^2}{V_b}\,.
 \ee 
 In the limit of small $Z_2$ symmetry breaking, one can use the ansatz $h= v_n \cos \phi(z)$ and $s= s_n \sin\phi(z)$ in the action, where $\phi(z) = \pi/4\,[1+\tanh(z/L_w)]$, and numerically solve for $c_1$ and $c_2$. This yields $c_1/c_2\approx 0.169$.\footnote{In Eq.~(46) of Ref.~\cite{Espinosa:2011eu}, the authors write a coefficient of $\sim 2.7$. This includes an extra factor of $16$ taken out of their expression for the potential barrier $V_b$. Thus our result agrees with theirs.} A similar argument made in \cite{Huber:2013kj} yields a coefficient of $\sim 1/8$, while Ref.\ \cite{Lewicki:2021pgr} obtained the value $\sim1/4$ by taking moments of the fields' equations of motion, but also showed that this approximation overestimated the width of the singlet wall by $\sim20 \%$, so we expect our estimate of $L_w$ to be valid up to a few percent. We find that this uncertainty affects the prediction for the BAU by a similar amount.

 \begin{figure}[]
    \centering
    \includegraphics[width=\linewidth]{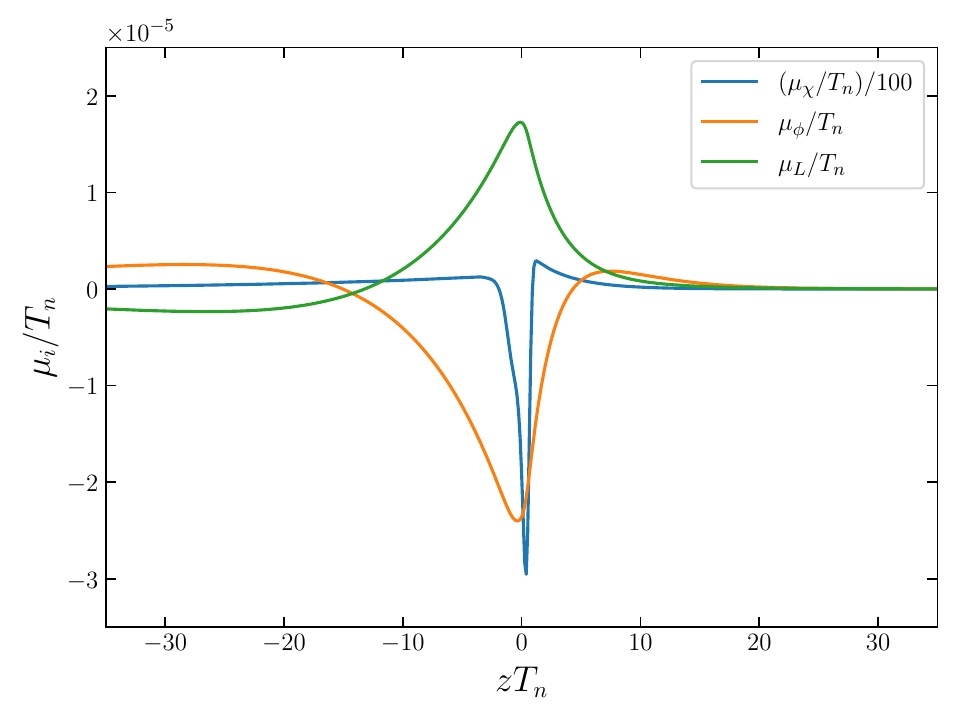}
    \caption{Solution to the transport equations for Model 1 presented in Table \ref{tab1}.}
    \label{fig:transport}
\end{figure}

\begin{figure*}[t]
    \centering
    \includegraphics[width=0.49\linewidth]{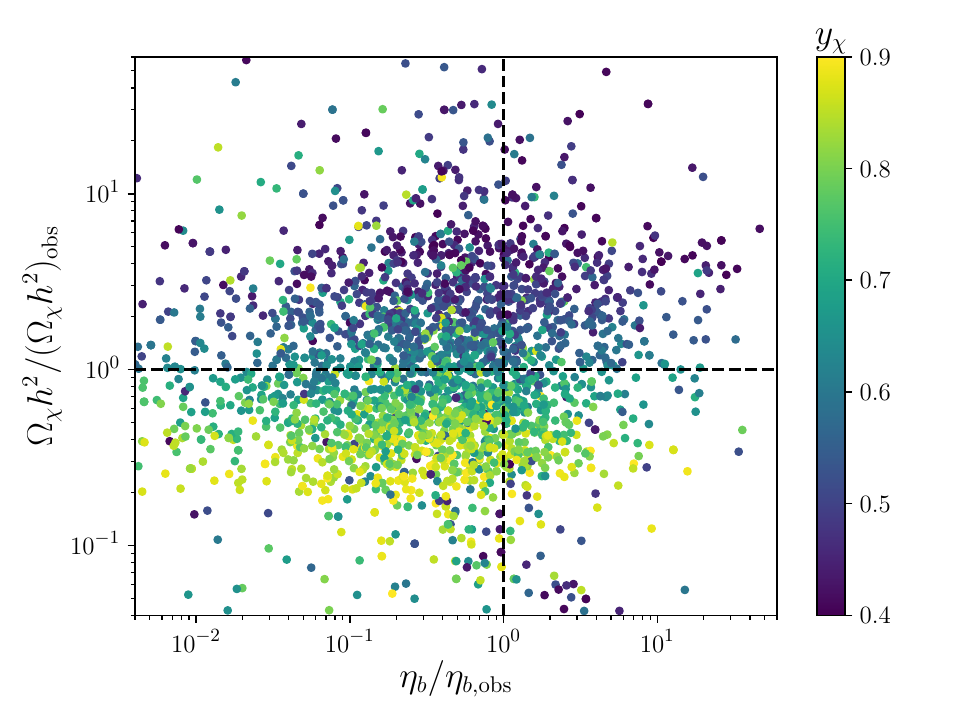}
    \includegraphics[width=0.49\linewidth]{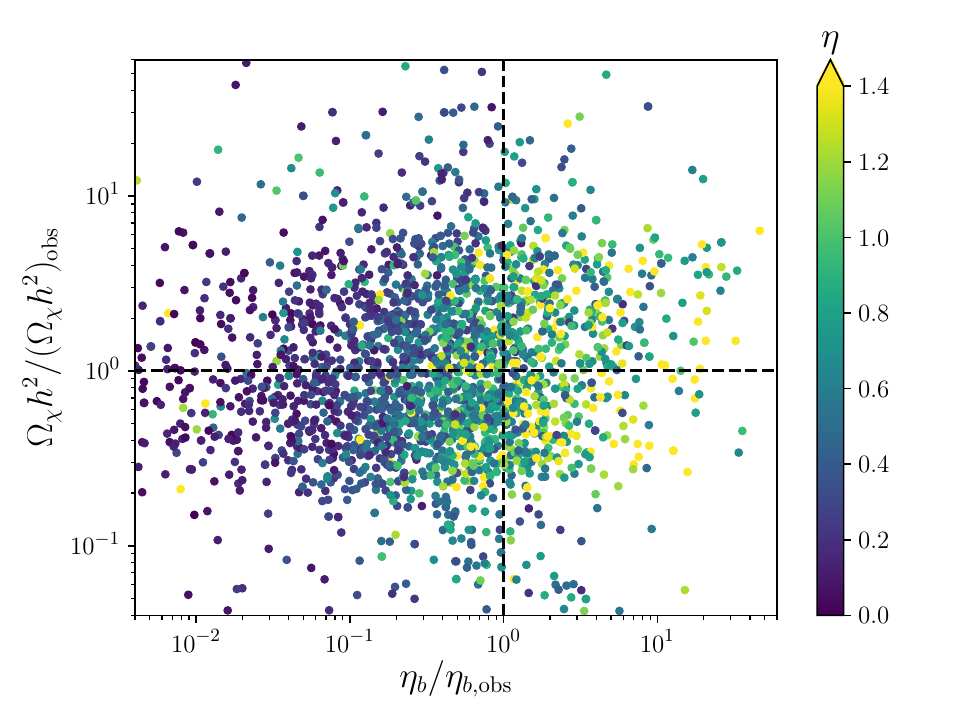}
    \caption{DM relic density and BAU for a sample of our Monte Carlo scan. Coloring of the dots indicates the values of the couplings $y_\chi$ (left) and $\eta$ (right) defined in Eq.\ (\ref{eq:chi-int}).}
    \label{fig:scan1}
\end{figure*}

\section{Electroweak Baryogenesis}
To estimate the resulting BAU in our model, we closely follow the analysis made by Ref.~\cite{Cline:2017qpe}, but with the revised transport equations presented in Ref.\ \cite{Cline:2020jre}, that are valid for high wall velocities. In the wall frame, the distribution function of a given species can be parametrized as
\be 
f=\frac{1}{e^{\beta\left[\gamma_w\left(E_w+v_w p_z\right)-\mu\right]} \pm 1}+\delta\! f\,,
\ee 
where the plus (minus) sign is for a fermion (boson), $\beta=1/T$, $\gamma_w=1/\sqrt{1-v_w^2}$, $\mu$ is a pseudochemical potential that parametrizes the local particle asymmetry and $\delta\! f$ is an unspecified function that satisfies
\be 
\int \mathrm{d}^3 p\, \delta\! f=0\,.
\ee

The time-independent Boltzmann equation is 
\be 
\left(v_g \partial_z+F \partial_{p_z}\right) f=\mathcal{C}[f]\,.
\ee 
We solve it by expanding $f$ to linear order in perturbations $\mu$ and $\delta$, and by taking the first two moments ($m=0,1$) of the resulting equation,
\be 
\left\langle\left(\frac{p_z}{E}\right)^{m} L\right\rangle=\left\langle\left(\frac{p_z}{E}\right)^{m}(\mathcal{S}+\delta \mathcal{C})\right\rangle.
\ee 
The moment average is defined as the phase space integral divided by a normalization factor,
\be 
\langle X\rangle \equiv \frac{1}{N_1} \int \mathrm{d}^3 p\, X,
\ee
where 
\be 
N_1 = \gamma_w \int d^3p\, {\partial f_{0,FD}\over \partial E} = - \gamma_w \frac{2\pi^3}{3} T^2. 
\label{N1eq}
\ee 
In Eq.\ (\ref{N1eq}), $f_{0,FD} = (e^{\beta E}-1)^{-1}$ is the equilibrium distribution function for a massless fermion in the fluid frame. Both bosons and fermions have the same definition for the normalization constant. This choice ensures that collision terms are equal for processes with equal rates.

We obtain a set of coupled equations for the pseudochemical
potential $\mu$ and the velocity dispersion $u$ of each species, defined as
\be 
u = \ev{\left({p_z\over E}\right)\delta\! f}\,.
\ee 
CP-odd components of these equations allow us to compute the resulting baryon asymmetry. The only relevant species are $\chi$, $\phi$ and the $SU(2)$ doublet $L_\tau$. Variation of the complex $\chi$ mass in the wall sources the asymmetry, which is then transferred to $L_\tau$ and the SM, principally via inverse decays of $\phi$. We  write the fluid equations in terms of the 2-vectors $w_i = (\mu_i,u_i)^T$,
\begin{align*}
&A_\chi w_\chi' - (m_\chi^2)' B_\chi w_\chi = S_\chi + \delta \mathcal{C}_\chi\,, \\
&A_\phi w_\phi'  = \delta \mathcal{C}_\phi\,,\\
&A_\tau w_{\tau}' = \delta \mathcal{C}_{\tau}\,,
\end{align*}
where 
\be
A=\begin{pmatrix}-D_1&1\\-D_2&R\end{pmatrix},\quad B=\begin{pmatrix}v_w\gamma_wQ_1&0\\v_w\gamma_wQ_2&\bar R\end{pmatrix}\,,
\ee
and the source term is $S_\chi=(S_1,S_2)^T$, with
\be 
S_{\ell}=-v_w\gamma_wh\big[(m^2\theta')'Q_\ell^{8o}-(m^2)'m^2\theta'Q_\ell^{9o}\big]\,.
\ee 
Primes indicate derivatives with respect to $z$, the direction transverse to the wall. The functions $D_\ell$, $Q_\ell$, $R$ and $\overline{R}$ are defined in Ref.\ \cite{Cline:2020jre}. In the source term, $m^2 = \abs{m_\chi(z)}^2$, $\theta=\arg(m_\chi(z))$ (cf.\ Eq.\ \ref{eq:chi_phase}) and $h=\pm 1$ is the helicity.\footnote{The CP-odd equation is obtained by subtracting the $h=+1$ equation from the $-1$ equation.} The collision vectors are $\delta\mathcal{C}_i=(K_{0}^i\,\delta C_{1}^i,\delta C_{2}^i)$, where
\begin{align*}\delta C_1^{\chi}&=2{\Gamma}_{\mathrm{hf}}\mu_{\chi}+2{\Gamma}_{d,\chi}(\mu_{\chi}+c\mu_{\tau}-c\mu_{\phi})\\
\delta C_1^{\phi}&={\Gamma}_{d,\phi}(\mu_{\phi}-\mu_{\tau}-c\mu_{\chi})+2{\Gamma}_{\times,\phi}(\mu_{\phi}-\mu_{\tau})\\
\delta C_1^{\tau}&={\Gamma}_{d,\tau}(\mu_{\tau}+c\mu_{\chi}-\mu_{\phi})+2{\Gamma}_{\times,\tau}(\mu_{\tau}-\mu_{\phi}),
\end{align*}
and $\delta C_2^i= -\Gamma_{tot}^i u_i-v_w K_0^i \delta C_1^i$. $K_0^i$ is a  factor that was introduced in \cite{Cline:2020jre} as a result of the particular normalization of the moment average we used. $\Gamma_d$ is the $\phi\to \chi L$ decay rate, $\Gamma_{hf}$ is the helicity-flipping rate due to $\chi\phi\to\chi\phi$ scattering, $\Gamma_\times$ is the $\phi\overline{L} \to \phi^* L$ scattering rate, $\Gamma_{tot}^i$ is the interaction rate, and $c$ is a helicity-flipping factor in $\phi$ decays. All of these parameters are derived in Appendices (B)-(D) of Ref.~\cite{Cline:2017qpe}.

The fluid equations can be combined into a simple linear system for the vector $U=(w_\chi, w_\phi, w_\tau)^T$,
\vskip-0.5cm
\be
    \qquad\qquad\qquad\qquad A U' =\Gamma U + S\,, \qquad\qquad\qquad\quad\ 
\ee

\noindent where $A=\mathrm{diag}(A_\chi,A_\phi,A_\tau)^T$, $\Gamma$ combines the collision terms and the $B$ matrix term, and $S=(S_\chi,0,0)^T$. This system can be solved with a relaxation algorithm, by imposing the boundary conditions $\mu_i =0$ at $z\to \pm \infty$. We used the Fortran routine TWPBPVPC \cite{CASH2005362} for this purpose.

With the solution for $\mu_\tau (z)$, one can compute the BAU using
\be 
\eta_{B}=\frac{405\Gamma_{\mathrm{sph}}}{4\pi^{2}v_{w}\gamma_{w}g_{*}T}\int dz\ \mu_{\tau}f_{\mathrm{sph}}e^{-45\Gamma_{\mathrm{sph}}|z|/4v_{w}\gamma_{w}},
\ee
where the sphaleron rate is $\Gamma_{\mathrm{sph}}=1.0\times 10^{-6}\, T$ and the factor
\be 
f_{\mathrm{sph}}=\min\left(1,\frac{2.4T}{\Gamma_{\mathrm{sph}}}e^{-40h(z)/T}\right)
\ee 
smoothly interpolates the sphaleron rate between the broken and unbroken phases. We then compared our result with the observed BAU, $\eta_{B0}=8.7\times 10^{-11}$.
Results of our MCMC scan (see Section \ref{sect:results}) for the correlation of BAU predictions versus dark matter relic density are shown in Fig.\ \ref{fig:scan1}. 
We are able to match the observed baryon asymmetry for reasonable values of the parameters, as we will further discuss in Section \ref{sect:results}.

\section{Dark matter relic density and GCE signal}

Two annihilation channels determine the relic density of dark matter, as shown in Fig.\ \ref{fig:ann_diagrams}. The first is annihilation into $\tau$ lepton doublets $L_\tau\bar{L}_\tau$ via $\phi$ exchange. The cross section is $p$-wave suppressed, and its thermal average is \cite{Cline:2017qpe}
\be \label{eq:cs_ll}
\langle\sigma v\rangle_{L_\tau\bar{L}_\tau }=\frac{y^4_\chi m_\chi\left(m_\chi^4+m_\phi^4\right) T}{4 \pi\left(m_\chi^2+m_\phi^2\right)^4}\,.
\ee 

The second channel is annihilation into SM left-handed fermions via Higgs-singlet exchange in the $s$ channel. In what follows we will only consider the dominant $b$ quark final state for the $s$ channel.  The cross section is 
 \begin{widetext}
\be\label{eq:sigma_b}
    \sigma(\mathfrak{s})_{b\bar{b} } = \frac{N_c y_b^2 \eta^2\theta_{hs}^2  (\mathfrak{s}-4m_b^2)^{3/2}}{32\pi \mathfrak{s} \sqrt{\mathfrak{s}-4m_\chi^2}} \frac{(m_h^2-m_s^2)^2 \left[\cos^2{\theta_\eta} (\mathfrak{s}-4m_\chi^2)+\sin^2{\theta_\eta} \mathfrak{s}\right]}{\left[(m_h^2-\mathfrak{s})^2+m_h^2\Gamma_h^2\right]\left[(m_s^2-\mathfrak{s})^2+m_s^2\Gamma_s^2\right]}\,
\ee
 \end{widetext}
where $N_c=3$, $y_b=\sqrt{2} m_b/v_0$  and $\mathfrak{s}=(p_1+p_2)^2$ is the Mandelstam variable. The decay widths are $\Gamma_h\approx 4.1$ MeV for the Higgs \cite{LHCHiggsCrossSectionWorkingGroup:2016ypw}  and 
\be 
\Gamma_s = \frac{\eta^2 m_s^2}{16 \pi } \left(1-\frac{4 m_\chi^2}{m_s^2}\right)^{3/2}
\ee 
for the gauge singlet, if $m_\chi < m_s/2$.  Despite being suppressed by the mixing angle $\theta_{hs}\ll1$, the cross section 
(\ref{eq:sigma_b}) is resonantly enhanced if $m_\chi$ is close to $ m_h/2$ or $ m_s/2$. This channel also has a velocity-independent ($s$-wave) contribution proportional to $\sin^2{\theta_\eta}$, as can be seen by taking the limit $\mathfrak{s}\to 4 m_\chi^2$ in the numerator. Due to the resonance, the partial wave expansion of the cross section is only accurate for small $v$. Instead, we computed the thermally averaged cross section using
\bea \label{eq:cs_bb}
\langle\sigma v\rangle &=& \frac{1}{8 m^4 T K_2^2(m / T)}\\&\times& \int_{4 m^2}^{\infty} d \mathfrak{s} \sqrt{\mathfrak{s}}\left(\mathfrak{s}-4 m^2\right) K_1(\sqrt{\mathfrak{s}} / T) \sigma(\mathfrak{s})\,,\nn
\eea 
where $K_i$ are the modified Bessel functions of the second kind.

Using the full annihilation cross section $\ev{\sigma v} = \ev{\sigma v}_{L_\tau\bar{L}_\tau}+\ev{\sigma v}_{b\bar{b}}$, we solved the following Boltzmann equation for the relic abundance $Y=n_{\chi}/s$,
\be 
\frac{d Y}{d x}=-\frac{x\left\langle\sigma v\right\rangle s}{H_{T=m_\chi}}\left( Y^2-Y_{eq}^2\right),
\ee 
where $x=m_\chi/T$, $H(T)=1.66\, g_*^{1/2}T^2/m_{Pl}$, $s =  (\pi^2/45) g_{*s}T^3$ and the nonrelativistic equilibrium abundance is
\be 
Y_{eq}(x) \approx 0.145 \frac{g_\chi}{g_{*S}}x^{3/2}e^{-x},\quad x\gg 1\,.
\ee 
The $\chi$ relic density is found using \cite{Kolb:1990vq}
\be 
\Omega_\chi h^2 = 2.82\times 10^8\ Y_\infty \left(\frac{m}{1\ \rm{GeV}}\right)\,.
\ee 
This result is to be compared with the observed dark matter relic density, $\Omega_{\rm DM}h^2=0.12$. In the following, we allow for the possibility that the $\chi$ particles need not account for all of dark matter in the universe, and could rather provide some fraction of the DM.

The $b$ quark annihilation channel could be responsible for the GCE signal, if it is dominant over the leptonic process, since annihilation into $\tau$ pairs has been shown to provide a poor fit to the data \cite{DiMauro:2021qcf}. It is therefore important to estimate cross sections for both channels in the galactic center, which requires knowledge of the DM velocity dispersion $\sigma_v^2$. Assuming dark matter follows a Maxwell-Boltzmann distribution, $f(v)\sim v^2 e^{-3 v^2/2\sigma_v^2}$, we can associate the velocity dispersion to an effective temperature $T= m_\chi\sigma_v^2/3$ to be used in Eqs.\ (\ref{eq:cs_ll}) and (\ref{eq:cs_bb}). The velocity dispersion is usually estimated from the virial mass of the galactic center, $\sigma_v^2 \sim GM/r\sim10^{-5}$, but it has been argued in Ref.~\cite{Johnson:2019hsm} that it could be up to $\mathcal{O}(10^{-2})$ due to `spikes' in the density profile of dark matter. We will therefore fix $\sigma_v^2=10^{-2}$ to assess the possibility that the $p$-wave $\tau$ channel might exceed the $b$ quark channel.

If 100\% of the DM annihilates and contributes to the GCE, current fits to the data require a thermal cross section of order $\ev{\sigma v}\sim 10^{-46}$ cm$^3$/s. But if $\chi$ only represents a fraction of the total DM, the annihilation cross section must be increased inversely proportional to the square of the $\chi$ relic density to compensate for the lower number of collisions in the galactic center. We will therefore consider the following effective cross section,
\be \label{eq:sigma_eff}
\ev{\widetilde{\sigma v}} \equiv \ev{\sigma v} \left(\frac{\Omega_\chi h^2}{\Omega_{\mathrm{DM}}h^2}\right)^2,
\ee 
to test whether our model can explain the origin of the GCE or not.

\section{Scan over model parameters}
\label{sect:results}

\begin{table*}
\centering
\begin{tabular}{ | c | c | c | c | c | c|c|c|c|c|c|c|c|} 
 \hline
 Model & $T_n$ & $v_n$ & $v_n/T_n$ & $\abs{s_n}$ & $s_0$ & $m_s$& $\theta_{hs}$ & $m_\chi$ & $\Omega_\chi h^2$ & $\eta_B/\eta_{B,\mathrm{obs}}$ & $\ev{\sigma v}_{b\bar b}/\ev{\sigma v}_{\tau\bar\tau}$ & $\Gamma_{\mathrm{inv}}$[MeV]\\\hline

 1 & 107.3& 205.9& 1.92& 156.5& 7.5& 97.6& $-$0.035& 49.3& 0.121& 0.95 & 8.5 & 0.30\\
 2 & 78.4& 232.9& 2.97& 188.6& 1.9& 104.3& $-$0.011& 64.2& 0.123 &1.02 & $4\times 10^{-4}$ & 0\\\hline
 Mean &  117.2 & 185.1 & 1.61&  125.2& 9.9& 151.3& 0.019 & 59.1& $-$ &$-$&$-$&$-$\\
 Std. dev. & 12.9 & 21.9& 0.41& 29.3 & 10.1&35.2 & 0.142&18.0& $-$ &$-$&$-$&$-$\\

\hline
\end{tabular}
\caption{Numerical results for the benchmark models of Table \ref{tab1} and statistical distribution among models that yield successful baryogenesis and DM abundance. Dimensionful quantities are in GeV, except for the Higgs invisible decay rate $\Gamma_{\mathrm{inv}}.$ Figures\ \ref{fig:scan1}, \ref{fig:gce} and \ref{fig:gammInv} illustrate the distribution of the last 4 quantities in our scan.}
\label{tab2}
\end{table*}

 To study the phenomenology of the phase transition, we performed a random scan over the parameter space of $\mu_s^2$, $\lambda_s$, $\lambda_{hs}$, $A_{hs}$, $A_3$, $M_\chi$, $\eta$ and $\theta_\eta$, with CosmoTransitions \cite{Wainwright:2011kj} for computing the bubble action and nucleation temperature of the EWPT. We used a Markov Chain Monte Carlo (MCMC) algorithm to focus on the regions of interest. Starting from a set of parameters that satisfies all constraints, other trials were generated by randomly varying all free parameters by small increments. Models where nucleation occurs were added to the Markov Chain with probability
\be 
P = \min\left(\frac{v_n/T_n}{1.1},1\right)
\ee 
as to favor strong first-order phase transitions with successful baryogenesis. Out of 150,000 models tested, we identified 23,797 which yield a FOPT with $v_n/T_n>1.1$. Fig.~\ref{fig:scan1} illustrates the resulting BAU and $\chi$ relic density in these models. The coloring of the dots highlights the correlation between $y_\chi$ and $\Omega_\chi h^2$ (left), as well as between the BAU and the coupling $\eta$ (right).

In our scan, 2,197 models satisfied the bounds
\begin{gather}
 \label{eq:models}\frac12 \le \frac{\eta_b}{\eta_{b,\rm obs}} \le 2\,,\\
\frac{\Omega_\chi h^2}{\Omega_{\rm DM} h^2} \leq 2\,.\nn 
\end{gather}
We found that, with small adjustments, these models can generally achieve the right BAU and not overproduce $\chi$ particles, which makes them good candidates for future phenomenological searches. We will therefore focus on those models when presenting the results of our scan. Two of those models have been identified in Table \ref{tab1} and will serve as benchmark models in our analysis. Important results for these benchmark models are presented in Table \ref{tab2}.

\begin{figure}[t]
    \centering
    \includegraphics[width=\linewidth]{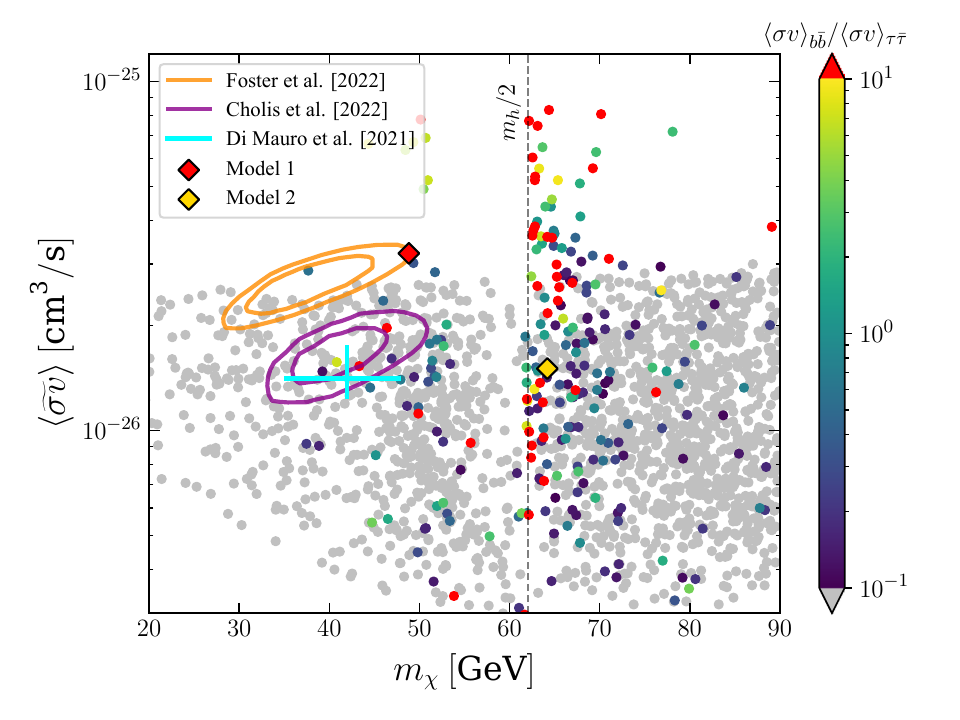}
    \caption{Effective DM annihilation cross section in the galactic center, assuming $\sigma_v^2\approx10^{-2}$. Coloring of the dots indicates the branching ratio to $b$ quarks versus $\tau$ leptons; the majority of models (grey dots) annihilate primarily into $\tau$. All models shown here satisfy the bounds (\ref{eq:models}) on the DM relic density and the BAU, hence the apparent cutoff on the vertical axis. The orange curves show $1$- and $2$-$\sigma$ best-fit regions to the GCE signal from Ref.\ \cite{Foster:2022nva} assuming a Higgs portal scenario. The magenta regions illustrate the best fit regions from Ref.\ \cite{Cholis:2021rpp} while the cyan cross is from Ref.\ \cite{DiMauro:2021qcf}, both assuming a $b$ quark annihilation channel. The two diamonds represent the benchmark models of Table\ \ref{tab1}; their branching ratios are given in Table\ \ref{tab2}.}
    \label{fig:gce}
\end{figure}
Fig.\ \ref{fig:gce} shows the effective DM annihilation cross section in the galactic center (cf.\ Eq.\  (\ref{eq:sigma_eff})), assuming a velocity dispersion $\sigma_v\approx10^{-2}$. Coloring of the dots illustrates the branching ratio of the two annihilation channels; the majority of models (grey dots) annihilate primarily into $\tau$ leptons. For comparison, we also present the best-fit regions to the GCE signal given Refs.\ \cite{Cholis:2021rpp}, \cite{Foster:2022nva} and \cite{DiMauro:2021qcf}. All fits indicate a preference for a DM mass $\sim30$--$50$ GeV and a cross section $\sim 10^{-26}$ cm$^3$/s. The small discrepancy between the best-fit cross sections is likely due to different assumptions made about the DM density profile in the galactic center and modeling of the GCE signal. Also shown are the two benchmark models, whose branching ratios are given in Table \ref{tab2}.

Model 1, in which $\chi$ annihilates mostly into $b$ quarks, lies just outside the $2\sigma$ region of the fit made in Ref.\ \cite{Foster:2022nva} to the GCE signal and could marginally explain the origin of the gamma rays. Model 2, like all models where $\chi$ primarily annihilates into leptons, does not fit the GCE data well. Ref.~\cite{DiMauro:2021qcf} showed that DM annihilation into muons or electrons could both explain the GCE, but the lepton channel is $p$-wave suppressed in our model, which makes it impossible to match the required cross section in the galactic center without overproducing $\chi$ particles, even if we assume the velocity dispersion is large. Couplings of $\chi$ to muons and electrons are also generally ruled out by experiments as we discuss in Appendix \ref{app:muon-electron}.

Models that fit the GCE signal are therefore dominated by the $b$ quark channel. This process is suppressed by the small mixing angle $\theta_{hs}$ between the singlet scalar and the Higgs boson, but it can be resonantly enhanced as explained below Eq.~(\ref{eq:sigma_b}). The imaginary part of the $\chi$-singlet coupling must also be large for the cross section to survive in the nonrelativistic limit of the galactic center. Given the $\chi$ mass favored by GCE data, models that explain the origin of the signal must generally satisfy
\begin{align*}
    m_\chi &\sim 30-50 \rm{\ GeV},\\
    m_s &\sim 2 m_\chi\,\\
    \sin\theta_\eta &\sim 1.
\end{align*}
Although Model 2 and the majority of the models in our scan do not provide a good explanation for the GCE, they can produce a large GW signal that could be detected in the near future. 

In Fig.\ \ref{fig:gw}, we showed the peak amplitude of the GW spectrum for the successful models in our scan, as well as the previously discussed PISC curves for future detectors BBO, DECIGO and LISA. Model 1 lies in the sensitivity region of BBO while Model 2, due to the strength of its phase transition, could produce a signal observable by all three experiments. 

\section{Collider and astrophysical constraints} 
In the previous section, we focused on the interplay between baryogenesis, gravitational waves, and the galactic center excess.  There are further constraints on the model from the Large Hadron Collider (LHC), Large Electron–Positron Collider (LEP), and direct/indirect dark matter detection, that we now discuss.  They remove an $\mathcal{O}(1)$ fraction of the previously allowed models, hence they represent complementary opportunites for  discovery.

\begin{figure}[t]
    \centering
    \includegraphics[width=\linewidth]{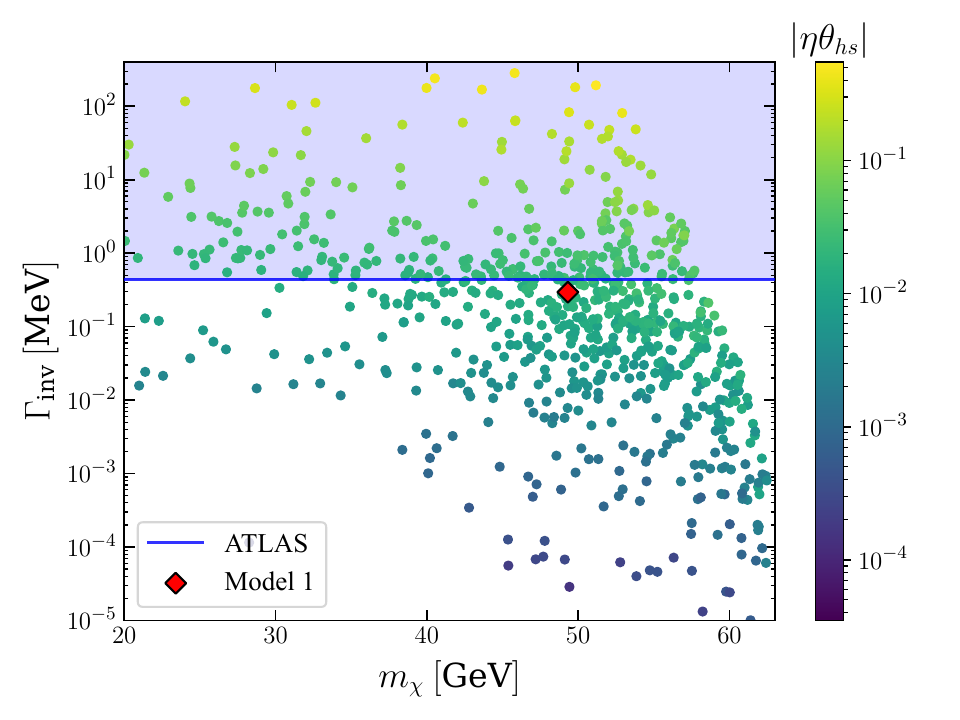}
    \caption{Higgs decay rate into $\chi$ pairs and upper limit set by constraints on invisible decay branching ratio \cite{ATLAS:2023tkt}. }
    \label{fig:gammInv}
\end{figure}

The Higgs invisible width is constrained to be at most 10.7\%  of its total width \cite{ATLAS:2023tkt}, which is 4.1\,MeV in the SM. Therefore, we require $\Gamma_{\mathrm{inv}}\lesssim0.44$ MeV.  To satisfy this for the $h\to ss$ channel, we impose that $m_s> m_h/2$ so that it is kinematically blocked.  If it were  allowed,  its rate would be $\Gamma\sim A_{hs}^2/(32 \pi m_h)$, which is ruled out for $A_{hs}\gtrsim $ few GeV, as we prefer for getting a strong EWPT.
If $m_\chi<m_h/2$, the decay $h\to\chi\bar\chi$ is allowed by Higgs-singlet mixing. Its width is
\be 
\Gamma_{\mathrm{inv}}= \frac{\eta^2 \theta_{hs}^2 m_h}{16 \pi}\sqrt{1-\frac{4 m_\chi^2}{m_h^2}}\left(1-\cos^2{\theta_\eta}\frac{4m_\chi^2}{m_h^2}\right).
\ee 
For a $\sim$50 GeV fermion, this leads to a limit $\abs{\eta\,\theta_{hs}}\lesssim 0.02$-$0.03$, depending on the phase of the singlet-DM coupling $\theta_\eta$,  becoming less stringent close to the Higgs resonance. Values of $\Gamma_{\mathrm{inv}}$ for models of our scan are illustrated in Fig.~\ref{fig:gammInv}. Model 1 lies just below the current bound and could thus be discovered in the near future.

\begin{figure}[t]
    \centering
    \includegraphics[width=\linewidth]{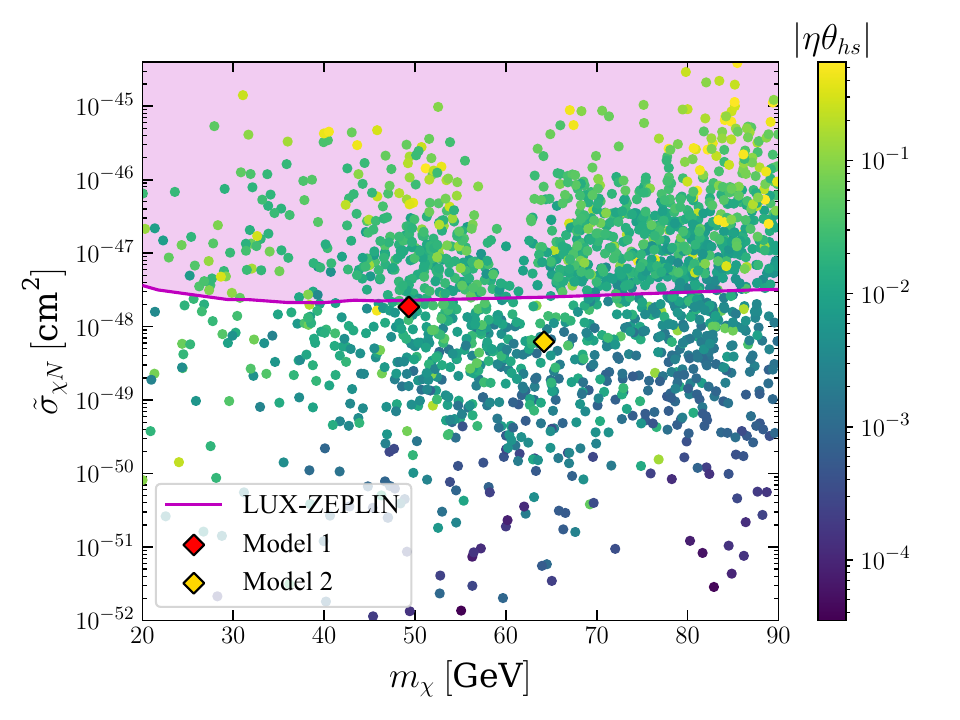}
    \caption{Direct detection cross section, rescaled by the $\chi$ relic density as explained in the text. The shaded region is excluded by LUX-ZEPLIN \cite{LZ:2024zvo}. }
    \label{fig:lz}
\end{figure}
Mixing of the singlet scalar and the Higgs boson induces a $\chi$-nucleon interaction, which could lead to the direct detection of dark matter. The interaction cross section is 
\be 
\sigma_{\chi \N}=\frac{1}{\pi}\left(\frac{\sin\theta_{hs}\eta\cos{\theta_\eta}y_\N\mu_{\chi \N}(m_h^2-m_s^2)}{m_h^2m_s^2}\right)^2,
\ee 
where $\mu_{\chi \N}$ is the reduced mass and $y_\N\approx 0.3\, m_\N/v_0$ is the Higgs coupling to nucleons. LUX-ZEPLIN (LZ) currently sets an upper limit of $\sigma_{\chi \N}\lesssim2\times 10^{-48}$ cm$^2$ \cite{LZ:2024zvo}, which corresponds to $\abs{\eta\theta_{hs}}\lesssim0.003$-$0.02$ depending on $m_s$ and the phase $\theta_\eta$. This limit is relaxed if $\chi$ does not constitute all of the dark matter. Fig.~\ref{fig:lz} illustrates the cross section rescaled by the $\chi$ relic density, $\tilde\sigma_{\chi N} \equiv \sigma_{\chi \N} (\Omega_\chi h^2/\Omega_{\mathrm{DM}}h^2)$, to compare the LZ bound with models where $\chi$ is a fractional component of the total dark matter.  Both benchmark models are slightly below the current limit.

\begin{figure}[t]
    \centering
    \includegraphics[width=\linewidth]{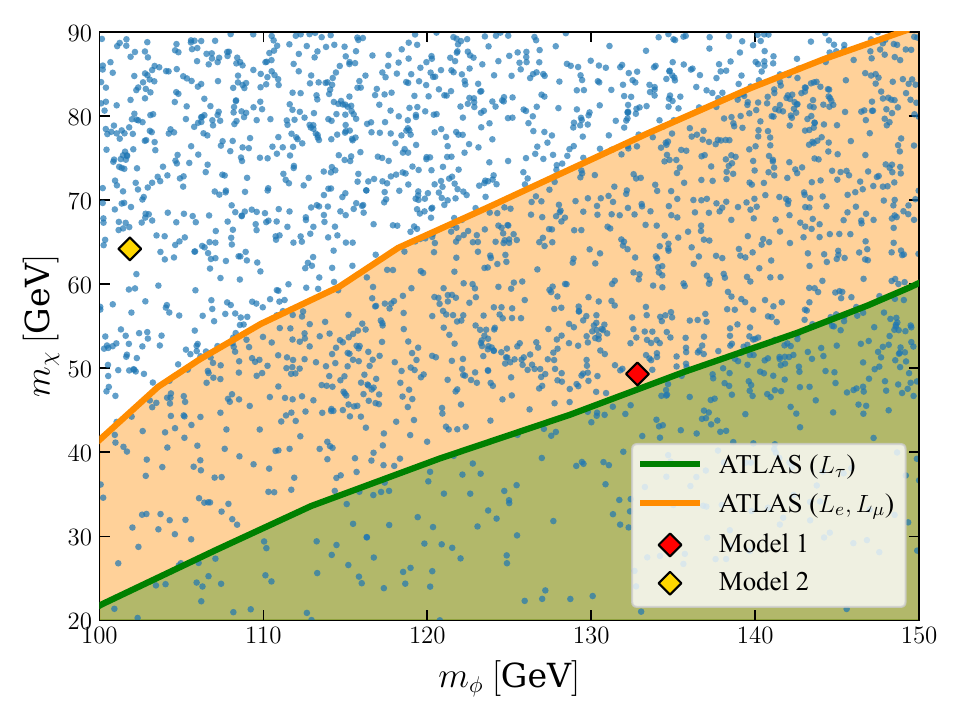}
    \caption{Distribution of $m_\chi$ and $m_\phi$ in our scan and comparison with bounds on Drell-Yan $\phi^\pm$ production at the LHC \cite{ATLAS:2018ojr,ATLAS:2024fub}, for scenarios where $\phi$ couples to $L_\tau$ (green) or $L_\mu,L_e$ (orange).} 
    \label{fig:slep_lim}
\end{figure}

At the LHC, Drell-Yan production of $\phi^\pm$ pairs and their subsequent decay into $\tau^\pm, \chi$ would mimic charged stau $\tilde\tau^\pm$ production and decay into tau-neutralino pairs in supersymmetric models, where $\chi$ and the neutralino would both appear as missing energy. Similar experiments at LEP have ruled out $m_\phi\lesssim 90$ GeV for our range of $m_\chi$ \cite{ALEPH:2003acj,DELPHI:2003uqw,L3:2003fyi,OPAL:2003nhx}. We therefore focus on $m_\phi>100$ GeV. We can directly apply limits set by ATLAS \cite{ATLAS:2018ojr,ATLAS:2024fub} on slepton production to our model since their cross sections are equal. Fig.~\ref{fig:slep_lim} illustrates constraints on stau production (green region) as well as on lighter slepton generations (orange) for comparison. Bounds on smuon and selectron production are more stringent and would typically require $m_\chi>m_\phi/2$.  Although the unshaded region of 
Fig.\ \ref{fig:slep_lim}  allows couplings to lower generation leptons, they are strongly constrained by electric dipole moment and anomalous magnetic moment experiments, as we discuss in Appendix \ref{app:muon-electron}.  For this reason, we consider couplings only to the $\tau$ in our scans.

At the one-loop level, the Yukawa coupling between $\phi$ and $L_\tau$ contributes to the effective $Z \bar{L}_\tau L_\tau$ coupling,  violating lepton flavor universality. In particular, the ratio of $Z\to e^+e^-$ to $Z\to \tau^+ \tau ^-$ decay widths is constrained by experiment \cite{ParticleDataGroup:2024cfk},
\begin{align*}
    {\Gamma(Z\to \tau^+\tau^-)\over \Gamma(Z\to e^+ e^-) }=1.0020\pm0.0032.
\end{align*}
In our model this ratio can be estimated as \cite{Khaw:2022qxh},\footnote{A similar argument can be made if $\phi$ couples to electrons only, which is marginally allowed, as explained in Appendix \ref{app:muon-electron}. One would then also have to consider deviations of the ratio $\Gamma(Z\to \mu^+\mu^-)/\Gamma(Z\to e^+e^-)$ from unity, which is only constrained to $\lesssim2.4\times10^{-3}$.}
\be 
{\Gamma(Z\to \tau^+\tau^-)\over \Gamma(Z\to e^+ e^-) }\approx 1+ \frac{2 g_L \delta g_L }{g_L^2+g_R^2}  \label{eq:zratio}
\ee 
where $g_L=-1/2+\sin^2\theta_W$, $g_R = \sin^2\theta_W$ are the $Z$ boson couplings in the SM and $\delta g_L$ is the contribution from new physics.  Ref.~\cite{Baker:2020vkh} 
shows that the one-loop contribution can be estimated as
\be 
\delta g_L \approx {y^2_\chi g_L\over 128\pi^2}
\ee 
up to a loop function $\sim\mathcal{O}(m_\chi^2/m_\phi^2)\lesssim1$. In our range of interest, $y_\chi\sim0.6$ and the ratio (\ref{eq:zratio}) differs from unity by $\sim3\times 10^{-4}$, which is well within the experimental uncertainty.

\begin{figure}[t]
    \centering
    \includegraphics[width=\linewidth]{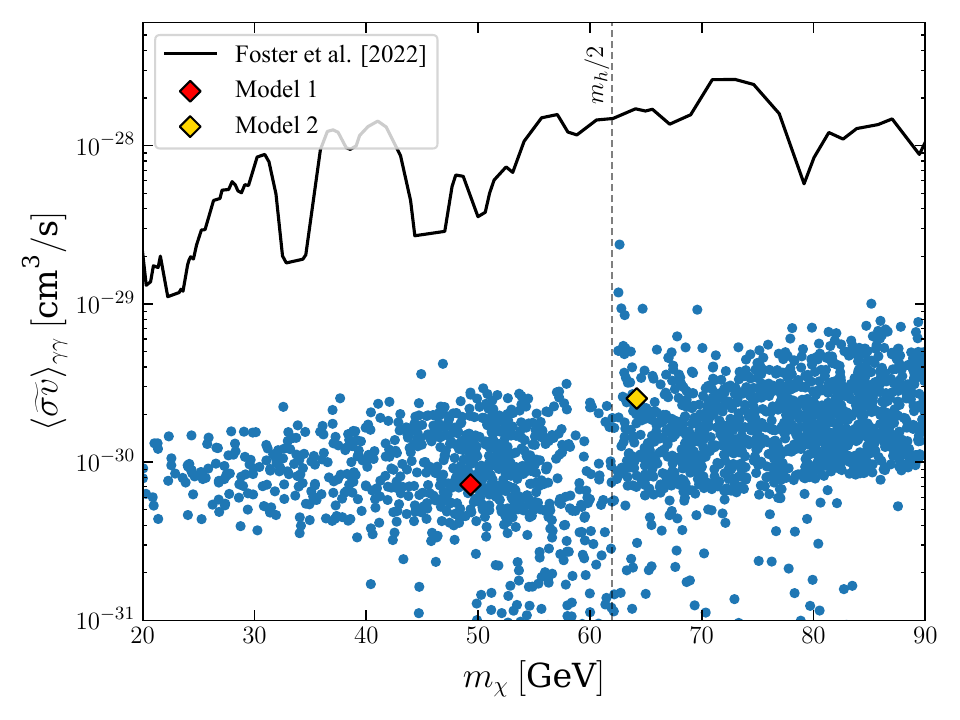}
    \caption{DM annihilation cross section into photons in the galactic center and upper limit from line search constraints presented in \cite{Foster:2022nva}. }
    \label{fig:line}
\end{figure}
DM annihilation into photons is also possible at the one-loop level by closing the $\tau$ or $b$ external lines in Fig.\ \ref{fig:ann_diagrams}. This generates a monochromatic signal at energy $E_\gamma=m_\chi$ coming from the galactic center, which is contrained by Fermi-LAT data \cite{Foster:2022nva}.
Ref.\ \cite{Garcia-Cely:2016hsk} computes this diagram in the nonrelativistic limit. For the virtual lepton contribution, the amplitude is
\bea
\mathcal{B}_{1} &=& \frac{\alpha y^2}{\sqrt{2}\pi} \left( \mathrm{Li}_2\left(- {m_\chi^2 \over m_\phi^2}\right) - \mathrm{Li}_2\left( {m_\chi^2 \over m_\phi^2}\right) \right) \nn\\ &\approx& - \frac{ \sqrt{2}\alpha y^2}{\pi} \left( m_\chi\over m_\phi \right)^2 + \mathcal{O}\left((m_\chi / m_\phi)^6\right),
\eea 
where $\mathrm{Li}_2(x)$ is the dilogarithm function.
The amplitude of the Higgs portal channel is 
\bea
\mathcal{B}_{2} &\approx& \frac{\sqrt{2} N_c \alpha y_f \eta\, \ss_{\theta_\eta}\theta_{hs}}{\pi}   \frac{m_\chi m_f(m_s^2-m_h^2)}{(m_h^2-4 m_\chi^2)(m_s^2-4 m_\chi^2)}\nn\\ &\times& \left(\pi - i \log\left(m_f^2\over 4 m_\chi^2\right)\right)^2\,,
\eea  
where we used $m_f\ll m_\chi$.
The annihilation cross section is then given by
\be 
\langle \sigma v\rangle_{\gamma \gamma } = \frac{\abs{\mathcal{B}_1 + \mathcal{B}_2}^2}{128 \pi m_\chi^2}\,.
\ee 
As before, we take the $\chi$ relic density into consideration by comparing the effective cross section defined in Eq.~\ref{eq:sigma_eff} with the limits set by Fermi-Lat. As shown in Fig.~\ref{fig:line}, the model predicts  $\langle \widetilde{ \sigma v} \rangle_{\gamma \gamma}\sim10^{-30}$ cm$^3$/s, which is about one order of magnitude below current bounds from monochromatic line searches \cite{Foster:2022nva}. 

\section{Conclusion}
We have revisited a model of the electroweak phase transition and dark matter that provides a very rich interplay of baryogenesis, gravitational waves, dark matter direct and indirect detection, 
including the galactic center gamma ray excess, and particle physics constraints.  With so many observables, the model is strongly constrained, but still viable in interesting regions of parameter space.  It predicts a fermionic dark matter candidate $\chi$ in the 50-100\,GeV range with Higgs portal couplings, a singlet scalar $s$ of mass $\sim 100$\,GeV that mixes with the Higgs, and an inert scalar doublet $\phi$ of mass $m_\phi\gtrsim 100$\,GeV.

In the original simpler version of the model, $Z_2$ symmetry was imposed on the singlet.  By relaxing that assumption in the present work, the phase transition can be strengthened, with advantages for GW production, and new interaction channels for the DM are opened, enhancing its detectability.  In particular, we were interested in the potential for explaining the long-standing excess of GeV gamma ray emission from the galactic center.  We find there is some tension to get a good fit to the GC excess while remaining consistent with other observables, but it is possible at the 2-$\sigma$ level, in models where $\chi\to b\bar b$ dominates over $\chi\to \tau^+\tau^-$.

We took advantage of advances that have been made to more accurately predicted the baryon asymmetry, including the tendency for faster bubble walls during the phase transition than was formerly thought.  For simplicity we assumed wall speeds close to the Jouguet velocity $\sim 0.6$.  The dual requirements of large enough baryon asymmetry and the correct DM relic density were realized without the need for fine tuning of parameters.

For gravitational waves, we verified that observable signals in future detectors is enhanced if the $Z_2$ breaking parameters $A_3$ or $A_{hs}$ in Eq.\ (\ref{V0eq})
are $\gtrsim 1\,$GeV, which is a modest requirement given that much higher values $\sim 100\,$GeV could be considered natural in the present context.  

In our setup, $\phi$ dominantly couples the dark matter to $\tau$ leptons, since couplings to lower generations are tightly constrained.
Our study of these couplings in Appendix \ref{app:muon-electron} indicates that there might be viable albeit finely-tuned models which dominantly couple to electrons, but they would require more careful computation of the two-loop contributions to the electron EDM, which is beyond the scope of the present work.

More generally, $\chi$ might interact with electrons and/or muons if their Yukawa couplings are at least ten times smaller than $\tau$, which would keep their EDMs and anomalous magnetic moments within experimental bounds. Most of our analysis would remain valid with this change. While it could lead to highly constrained flavor-changing neutral currents at the one-loop level, it also opens the possibility of scotogenic neutrino masses \cite{Ma:2006km} if one includes the lepton number-violating interaction
\be 
\sim\lambda'(\phi^\dagger H)^2 + \mathrm{h.c.}
\ee 

Extending our model to include multiple Majorana fermions $\chi_m$ and/or inert doublet $\phi_n$, this interaction would lead to one-loop neutrino mass terms of order
\be 
m_{ij}\simeq \sum_{m,n} \frac{y_{imn} y_{jmn}}{16 \pi^2} \frac{\lambda' v_0^2 m_{\chi_{m}} }{ m_{\phi_n}^2}\,,
\ee 
where $y_{imn}$ is the Yukawa coupling of $L_i$ to $\chi_m$ and $\phi_n$.\footnote{With a single fermion $\chi$ and a single scalar doublet $\phi$, the mass matrix $m_{ij}\propto y_iy_j$ would have two zero eigenvalues.} Given that $y_\chi\sim 0.6$ is needed for the DM relic density and baryogenesis, a coupling $\lambda_{\phi H}\sim10^{-10}$ could produce neutrino masses of order $m_\nu\sim0.05$ eV. However, one would have to consider the bounds on muon and electron couplings listed in this work as well as constraints on neutrino mixing and lepton flavor universality to build a complete physical model. A similar setup was studied in Refs.~\cite{Abada:2018zra,Fujiwara:2020unw}.

In summary, the model presents opportunities for future GW detection that could be corroborated by direct detection of dark matter of mass
$\sim 50\,$GeV, LHC searches for stau-like pair production,
invisible decays of the Higgs, and it can explain the galactic center excess within $\sim 2\,\sigma$ confidence level, while providing for successful baryogenesis.

\bigskip

{\bf Acknowledgments.}  We thank K.\ Kainulainen, B.\ Laurent and D.\ Tucker-Smith for helpful correspondence. This work was supported by the Natural Sciences and Engineering Research Council (Canada) and the Fonds de recherche Nature et technologies (Qu\'ebec). 

\appendix
\section{Effective Potential and Mass Eigenvalues}
\label{app:masseigenvalues}
When the present model was first proposed in Ref.\ \cite{Cline:2017qpe}, the authors only considered the tree-level potential Eq.~(\ref{eq:pot}) and used the high-temperature approximation for thermal corrections. Our study improves on the original paper by including the Coleman-Weinberg (CW) potential as well as counterterms and the full thermal potential. Counterterms and  renormalization conditions are discussed in Appendix \ref{app:counterterms}. 

Denoting the two relevant scalar fields as $\Phi=(h,s)$, the CW potential is
\bea \label{eq:CW}
V_{\mathrm{CW}}(\Phi,T)&=&\sum_i(-1)^F \frac{g_i m_i^4(\Phi,T)}{64 \pi^2}\nn\\
&\times&\left(\log \left(\frac{m_i^2(\Phi,T)}{Q^2}\right)-C_i\right),
\eea
where $F=1\ (0)$ for fermions (bosons) and $g_i=1/2/3$ for scalars, fermions and vectors respectively. In the $\overline{\mathrm{MS}}$ renormalization scheme, $C_i=3/2$ for scalars, fermions and longitudinal polarizations of gauge bosons, and $C_i=1/2$ for transverse gauge bosons. We set the normalization scale to $Q^2=v_0^2$.

The thermal corrections to the scalar potential are given by
\be \label{eq:VT}
V_T\left(\Phi, T\right)=\frac{T^4}{2 \pi^2} \sum_i(-1)^F g_i J_{B / F}\left(\frac{m_i^2\left(\Phi_i,T\right)}{T^2}\right),
\ee 
where the thermal functions are
\be 
J_{B / F}\left(z^2\right)=\int_0^{\infty}\!\!\!\!\! dx\, x^2 \log \left[1-(-1)^F \exp \left(-\sqrt{x^2+z^2}\right)\right].
\ee 
In Eqs.~(\ref{eq:CW}) and (\ref{eq:VT}), $m_i(\Phi,T)$ are the field-dependent mass eigenvalues of all species. Following Parwani (daisy) resummation, we also include the leading-order temperature dependence of scalar mass eigenvalues in the one-loop and thermal potentials. 

We next present the relevant field-dependent masses. The Higgs-singlet scalar mass matrix is
\begin{widetext}
\begin{align}
\mathcal{M}^2 &= \begin{pmatrix}
		-\mu_h^2 + 3 \lambda_h h^2 + \frac{1}{2} \lambda_{hs} s^2 - A_{hs} s & \lambda_{hs} hs-A_{hs} h\\
		\lambda_{hs} hs-A_{hs} h & -\mu_s^2 + 3 \lambda_s s^2 +\frac{1}{2}\lambda_{hs}h^2 -2 A_3 s
	\end{pmatrix}\nn\\
&+ \frac{1}{48} \begin{pmatrix}
	9g^2+3g'^2+12 y_t^2+24\lambda_h+2\lambda_{hs} & 0\\
	0& 8\lambda_{hs}+12\lambda_s+4\eta^2
\end{pmatrix}T^2,\label{eq:m2scalar}
\end{align}
\end{widetext}
where $g$ and $g'$ are respectively the $SU(2)$ and $U(1)$ gauge couplings, and $y_t=\sqrt{2}m_t/v_0$ is the top quark Yukawa coupling. The field-dependent mass eigenvalues are obtained by diagonalizing $\mathcal{M}^2$,
\be 
m_h^2,m_s^2 = \frac{\mathcal{M}_{hh}^2 + \mathcal{M}_{ss}^2\pm \sqrt{(\mathcal{M}_{hh}^2+\mathcal{M}_{ss}^2)^2-4\Delta}}{2}, 
\ee 
where $\Delta=\mathcal{M}_{hh}^2\mathcal{M}_{ss}^2-(\mathcal{M}_{hs}^2)^2$. Vaccuum stability requires $\Delta>0$ in the true and metastable vacua.

In unitary gauge, Goldstone bosons do not enter the tree-level potential, but their contributions to the CW and thermal potential must be included. The mass eigenvalue is
\bea
	m_{GB}^2&=&-\mu_h^2+\lambda_h h^2-A_{hs}s+\sfrac{1}{2}\lambda_{hs}s^2\\ &+&
\sfrac{1}{48}\left[9 g^2+3 g^{\prime 2}+12 y_t^2+24 \lambda_h+2\lambda_{hs}\right] T^2\nn
\eea
At $T=0$, Goldstone bosons become massless in the true vacuum. This results in a well-known issue of making the counterterms singular, arising from the Higgs self-energy evaluated at zero external momentum rather than on shell in the CW potential \cite{Casas:1994us,Cline:1996mga}. The IR divergence is therefore unphysical and can be avoided by following the procedure described in Ref.\ \cite{Elias-Miro:2014pca}.

We also include the contribution from massive gauge bosons in the thermal potential. In the $(W^\pm,W^3,B)$ basis,
the mass matrix is
\begin{widetext}
\begin{align}
\mathcal{M}_{\rm gauge}^2 &= 
	\begin{pmatrix}
g^2 & & & \\
& g^2 & & \\
& & g^2 & g g^{\prime} \\
& & g g^{\prime} & g^{\prime 2}
	\end{pmatrix} \frac{h^2}{4} + \begin{pmatrix}
	g^2 & & \\
	& g^2 & & \\
	& & g^2 & \\
	& & & g^{\prime 2}
\end{pmatrix}T^2\ \begin{cases}2, & \text { longitudinal; } \\
0, & \text { transverse. }\end{cases}
\end{align}
\end{widetext}
The inert doublet $\phi$ mass does not depend on $(h,s)$, and it does not acquire a VEV during the EWPT. Hence it does not impact the dynamics of the phase transition and can be ignored with respect to the effective potential for $h$ and $s$.  (For simplicity we have omitted the $|H|^2|\phi|^2$ coupling).

In the fermionic sector, we include the contributions from the top quark, $m_t=y_t h/\sqrt{2}$, and the Majorana fermion,
\begin{align} \label{eq:chimass}
	m^2_\chi& = (M_\chi+\eta \cos \theta_\eta\, s)^2 + (\eta \sin \theta_\eta\, s)^2\nn\\
	& = M_\chi^2 + 2 M_\chi \eta \cos\theta_\eta\, s+ \eta^2 s^2 .
\end{align}
Recall that $M_\chi$ is the bare mass term in Eq.~(\ref{eq:chi-int}). Lighter SM fermions contribute negligibly to the one-loop potential.

\section{Counterterms}
\label{app:counterterms}

At $T=0$, the CW potential shifts observables away from their tree-level values, in particular the masses and VEVs of the scalar fields. To counteract this, we introduce the counterterm potential,
\bea
\delta V &=&-\sfrac{1}{2} \delta \mu_h^2 h^2+\sfrac{1}{4} \delta \lambda_h h^4-\sfrac{1}{2} \delta \mu_s^2 s^2+\sfrac{1}{4} \delta \lambda_s s^4\\&+&\sfrac{1}{4} \delta \lambda_{hs} h^2 s^2 - \sfrac{1}{2} \delta A_{hs} s h^2+\delta A_1^3 s-\sfrac{1}{3} \delta A_3 s^3.\nn
\eea
 The eight parameters in $\delta V$ are chosen so that tree-level relations are preserved at $T=0$. The full zero-temperature, one-loop potential is $\tilde V\equiv V_{CW}+\delta V$. Although the tree-level potential does not include a tadpole term for $s$, such a term can arise from $V_{CW}$, necessitating the tadpole counterterm $\delta A_1^3 s$.

The renormalization conditions we use are essentially the same as those presented in the appendix of Ref.~\cite{Espinosa:2011ax}. 
 In the following we denote the electroweak symmetric vacuum by $S\equiv(h,s)=(0,s_1)$ and the broken vacuum by $B\equiv(h,s)=(v_0,s_0)$. We impose these conditions to fix the counterterms:

\begin{itemize}
	\item The Higgs and singlet VEVs are unaffected by $\tilde V$, in both the broken and unbroken vacua.
	\be 
	\left.\pdv{\tilde V}{h}\right|_{B} = \left.\pdv{\tilde V}{s}\right|_{B} = \left.\pdv{\tilde V}{s}\right|_{S} = 0
	\ee
	\item The scalar mass matrix elements are equal to their tree-level values in the broken vacuum.
	\be 
	\left.\pdv[2]{\tilde V}{h}\right|_{B} = \left.\pdv[2]{\tilde V}{s}\right|_{B} = \left.\pdv{\tilde V}{h}{s}\right|_{B} = 0
	\ee
	\item The cubic coupling is equal to its tree-level value in the true minimum.
	\be 
	\left.\pdv[3]{\tilde V}{s}\right|_{B}=0
	\ee 
	\item The energy difference between the two vacua is the same as at tree level.
	\be 
	\left.\tilde{V}\right|_B - \left.\tilde{V}\right|_S = 0
	\ee
\end{itemize}
Solving for the counterterms yields 
\begin{widetext}
\begingroup 
\allowdisplaybreaks
\begin{align} 
\label{dmuheq} \delta\mu_h^2 &= \frac{1}{2v_0^2} \bigg\lbrace\Big[(3-5x_s^2)\partial_{\tilde h} - (1-x_s^2)\partial^2_{\tilde h} - x_s(2+3x_s)\partial^2_{\tilde h \tilde s} \Big]\tilde V|_B \nn\\
&\left. + 2x_s^2 \left[3\partial_{\tilde s}+\partial^2_{\tilde s}+\frac{1}{6}\partial^3_{\tilde s}\right]\tilde V|_B + 2x_s^2(\partial_{\tilde s}\tilde V|_S+4\Delta \tilde V_{BS}) \right\rbrace\\
\label{dlheq}
\delta \lambda_h&=\frac{1}{2v_0^4}\left(\partial_{\tilde h}-\partial^2_{\tilde h}\right)\tilde V|_B\ \\ 
\label{dmuseq}\delta \mu_s^2 & = \frac{1}{2\Delta s^2} \bigg\lbrace (1-3x_s^2)\Big[\left(5\partial_{\tilde h} - \partial_{\tilde h}^2\right)\tilde V|_B -8 \Delta \tilde V_{BS} \Big]-2(1-6x_s^2)\partial_{\tilde s}\tilde V|_S \nn \\
& +\bigg[ 3 (1-2x^2_s)(\partial_{\tilde h \tilde s}^2- 2\partial_{\tilde s})-\frac{1}{3}(1+6x_s+6x_s^2)\partial_{\tilde s}^3  \bigg] \tilde V|_B \bigg\rbrace\\
\label{dlseq}\delta \lambda_s &= \frac{1}{2\Delta s^4} \left\lbrace\left((5\partial_{\tilde h} - \partial_{\tilde h}^2-4\partial_{\tilde s}+2 \partial^2_{\tilde h \tilde s}+\frac{2}{3}\partial_{\tilde s}^3\right) \tilde V|_B -4 \partial_{\tilde s}\tilde V|_S -8 \Delta \tilde V_{BS} \right\rbrace\\
\label{dlhseq}\delta \lambda_{hs}& = \frac{1}{v_0^2 \Delta s^2} \left\lbrace\left(5\partial_{\tilde h}-\partial^2_{\tilde h}-6\partial_{\tilde s}-2\partial^2_{\tilde s}+3\partial^2_{\tilde h \tilde s} - \frac{1}{3}\partial^3_{\tilde s}\right)\tilde V|_B -2 \partial_{\tilde s}\tilde V|_S -8 \Delta \tilde V_{BS} \right\rbrace\\
\label{dahseq}\delta A_{hs} &= \frac{1}{v_0^2 \Delta s}\bigg\lbrace \left[x_s (5 \partial_{\tilde h}-\partial^2_{\tilde h}-6\partial_{\tilde s}-2\partial_{\tilde s}^2 -\frac{1}{3}\partial^3_{\tilde s} ) +(1+3x_s)\partial^2_{\tilde h \tilde s} \right]\tilde V|_B\nn\\ &-2x_s \left( \partial_{\tilde s} \tilde V|_S + 4 \Delta \tilde V_{BS}\right)\bigg\rbrace\\
\label{da1eq}\delta A_1^3 & =\nn \frac{1}{2\Delta s} \bigg\lbrace -2x_s\left[4(1-x^2_s)\Delta \tilde V_{BS}+ (1-2x^2_s) \partial_{\tilde s}\tilde V|_S\right]\\ &\nn+(1+x_s)\left[x_s(1-x_s)(5\partial_{\tilde h} - \partial^2_{\tilde h})-\frac{1}{3}x_s(1+2x_s)\partial^3_{\tilde s}\right.\\&-(1+2x_s-2x^2_s)(2\partial_{\tilde s}-\partial^2_{\tilde h \tilde s})\bigg]\tilde V|_B\bigg\rbrace\\
\label{da3eq}\delta A_3 & = \frac{3}{2\Delta s^3} \bigg\lbrace \left[x_s(5\partial_{\tilde h}-\partial_{\tilde h}^2-4\partial_{\tilde s}+2\partial^2_{\tilde h \tilde s})+\frac{1}{3}(1+2x_s)\partial^3_{\tilde s} \right]\tilde V|_B \nn\\ & -4x_s(\partial_{\tilde s}\tilde V|_S + 2 \Delta \tilde V_{BS})\bigg\rbrace
\end{align} 
\endgroup
\end{widetext}
where $\Delta s = s_1-s_0$, $x_s = s_0/\Delta s$, $\tilde h \equiv h/v_0$, $\tilde s \equiv s/\Delta s$ and $\Delta \tilde V_{BS} = \tilde V|_B-\tilde V|_S$.
Some coefficients in Eqs.\ (\ref{dlheq}), \ref{dlseq},\ref{dahseq}-\ref{da3eq}) do not match the corresponding values in Eqs.\ (A.12), (A.14), (A.9), (A.11) and (A.10) of Ref.\ \cite{Espinosa:2011ax} despite using the same renormalization conditions. 
We have carefully checked our calculations and do not find any error on our part.

\section{Beyond the $\tau$ coupling}
\label{app:muon-electron}

\begin{figure}[t]
    \centering
    \raisebox{0.8cm}{\includegraphics[scale=0.5,page=4]{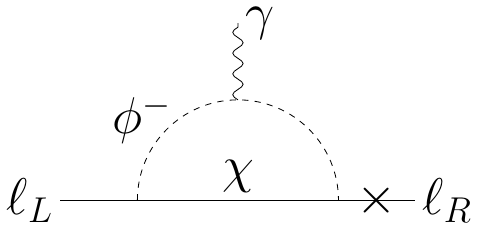}}
    \includegraphics[scale=0.5,page=3]{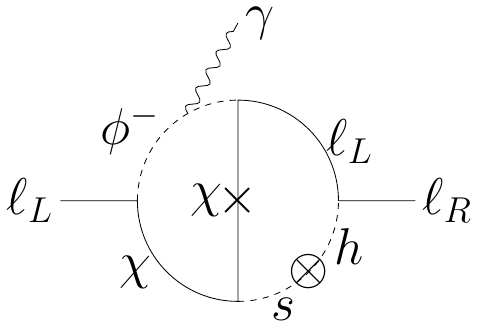}
    \caption{Examples of Feynman loop diagrams contributing to the lepton anomalous magnetic moment (left) and electric dipole moment (right). Crosses indicate fermion mass insertions and a cross within a circle represents the Higgs-singlet mixing. For the electric dipole moment, there is a similar diagram where the photon is attached to the lepton inside the loop.}
    \label{fig:g-2edm}
\end{figure}

In this work we have focused on the scenario where the Majorana fermion dark matter candidate $\chi$ only interacts with $\tau$ leptons at tree level. Here we discuss the possibility and associated constraints of coupling $\chi$ to other lepton flavors.

The coupling of the dark sector to a charged SM lepton $\ell_i$ contributes to its anomalous magnetic dipole moment $a_i=(g_i-2)/2$ via the loop diagram on the left in Fig.~\ref{fig:g-2edm}. The induced shift in $a_i$ is
\begin{align}
\abs{\delta a_i} &= \frac{y^2_\chi }{16 \pi^2 } \frac{m_\ell^2}{m_\phi^2}\  f\!\left(\frac{m_\ell^2}{m_\phi^2}\right), \label{eq:g-2}
\\ \nn f(x) &= \frac{2}{(1-x)^4} \left(1 - 6x +3x^2+2x^3-6x^2\ln{x} \right).
\end{align}

The electron magnetic moment is measured with a precision of 0.13 ppt ($1.3\times 10^{-13}$) \cite{Fan:2022eto}. This result differs by $2.1\sigma$ and $3.9\sigma$ from theoretical predictions obtained from measurements of the fine-structure constant in Rb \cite{Morel:2020dww} and Cs \cite{Parker:2018vye} atoms respectively. However, those predictions are in tension at the level of 1 ppt ($10^{-12}$) with each other, which dominates over the experimental precision. Our dark sector could therefore contribute to the electron magnetic moment up to $\abs{\delta a_e}\sim 10^{-12}$ without being ruled out. The value of $\abs{\delta a_e}$ is however suppressed by the small electron mass, so that contributions from new physics are generally small. 

The muon magnetic moment, while less severely constrained, is more sensitive to physics beyond the standard model (BSM). Until recently, there was a discrepancy of $\mathcal{O}(10^{-9})$ between the measured and predicted values of $a_\mu$ \cite{Muong-2:2021ojo}. Following recent theoretical calculations \cite{Aliberti:2025beg} and experimental results from the Muon $g-2$ collaboration \cite{Muong-2:2025xyk}, this tension is now gone and only allows for a contribution $\abs{\delta a_\mu}\lesssim 2\times 10^{-10}$ from a dark sector. 

For parameters of interest, Eq.~(\ref{eq:g-2}) indicates $\abs{\delta a_\mu}\sim 10^{-9}$--$10^{-8}$. This rules out muons as the lepton flavor that couples dominantly to the dark sector. If we consider coupling to electrons instead, we find $\abs{\delta a_e} \sim 10^{-13}$ for our benchmark models, which is of the same order of magnitude as the current experimental precision and smaller than the discrepancy between theoretical predictions. For tau leptons, we find $\abs{\delta a_\tau}\sim 10^{-6}$, which is one order of magnitude below the predicted sensitivity of upcoming experiments \cite{Nakai:2025dmp}. 

Our model includes a new source of CP violation coming from the complex $\chi$-singlet coupling and, by extension, the complex $\chi$ mass whose phase is given by Eq.~(\ref{eq:chi_phase}). CP violation in BSM models can induce an  electric dipole moment (EDM) for the electron, which is highly constrained \cite{ACME:2018yjb},
\be 
d_e < 4.1\times10^{-30} \ e\ \mathrm{cm}\,.
\ee 

To estimate the magnitude of the EDM, it is useful to perform a chiral rotation $\chi\to e^{-i \gamma_5\theta_{\chi0}/2}\chi$, where $\theta_{\chi0}$ is the argument of the complex $\chi$ mass in the true vacuum where $\ev{s}=s_0$. Then the only effective CP-violating phase  comes from the $\chi$-singlet vertex coupling with complex argument
\be 
\theta_{\mathrm{CP}} = \theta_\eta-\theta_{\chi0}.
\ee 

In BSM models where new physics does not interact with SM singlet leptons $\ell_R$, an EDM arises at the two-loop level only \cite{Ardu:2025awk}. In our model, the diagram that induces an EDM for the charged lepton $\ell_i$ that couples to $\chi$ is illustrated in Fig.~\ref{fig:g-2edm} (right). We estimate the amplitude of the EDM to be
\be 
d_i \simeq \frac{e  y_\ell y_\chi^2 \theta_{hs} \eta \sin{\theta_{\mathrm{CP}}} m_\chi }{(16\pi^2 \Lambda)^2} \,,
\ee 
where the lepton coupling to the Higgs boson is $y_\ell=\sqrt{2} m_\ell/v_0$ and we assumed that the loop is dominated by
momenta of order $\Lambda=\max(m_s,m_h,m_\phi)$.\footnote{We ignored a partial cancellation $\propto (m_s^2-m_h^2)$ between the two mass eigenstates that run in the loop, which might suppress the EDM if the states are nearly degenerate.} Despite being suppressed by the Higgs-singlet mixing angle, the EDM resulting from this diagram would generally  be $\gtrsim 10^{-29}\,e$ cm if $\chi$ couples to the electron, which is ruled out. Although some models in our scan predict an EDM allowed by current bounds, they generally have a weaker Yukawa coupling $y_\chi$
 or a larger $m_\phi$, which is difficult to reconcile with the observed DM relic density. Model 2 of Table \ref{tab1} is a notable exception.\footnote{Our estimate predicts $d_e\sim 10^{-30}\,e$ cm for Model 2. A precise calculation of the two-loop integral would be needed to better assess the possibility of coupling this model to the electron.} The tau lepton EDM predicted by our model is $d_\tau\sim 10^{-25}\ e$ cm, well below the sensitivity of upcoming experiments \cite{Nakai:2025dmp}. Therefore, dominantly coupling $\tau$ to $\chi$ is the most viable option.

\bibliography{ref}
\bibliographystyle{utphys}
\end{document}